\documentclass[a4paper,prb,twocolumn,groupedaddress,showpacs]{revtex4}

\usepackage[utf8]{inputenc}
\usepackage{amssymb}
\usepackage{amsmath}
\usepackage{color}
\usepackage{graphics, graphicx}
\usepackage{bbold}
\usepackage{psfrag}
\usepackage{mathcomp}
\usepackage{subfigure}
\usepackage{booktabs}
\usepackage{blindtext}
\usepackage{float}
\usepackage{import}
\usepackage{sansmath}

\usepackage[pdftitle={},
            pdfsubject={},
            pdfauthor={Steffen Backes},
            ]{hyperref}

\usepackage{color}
\usepackage{soul}

\newcommand{\ba}{BaFe$_2$As$_2$}
\newcommand{\kf}{KFe$_2$As$_2$}
\newcommand{\cs}{CsFe$_2$As$_2$}
\newcommand{\rb}{RbFe$_2$As$_2$}
\newcommand{\fr}{FrFe$_2$As$_2$}

\begin{document}

\author{Steffen Backes}
\email{backes@itp.uni-frankfurt.de}
\affiliation{Institut f\"ur Theoretische Physik, Goethe-Universit\"at Frankfurt, Max-von-Laue-Str. 1, 60438 Frankfurt am Main, Germany}
\author{Harald O. Jeschke}
\affiliation{Institut f\"ur Theoretische Physik, Goethe-Universit\"at Frankfurt, Max-von-Laue-Str. 1, 60438 Frankfurt am Main, Germany}
\author{Roser Valent\'{\i}}
\affiliation{Institut f\"ur Theoretische Physik, Goethe-Universit\"at Frankfurt, Max-von-Laue-Str. 1, 60438 Frankfurt am Main, Germany}

\date{\today}


\
\vspace{0.5cm}
\

\title{Microscopic nature of correlations in multi-orbital $A$Fe$_2$As$_2$ ($A=$K, Rb, Cs):
 Hund's coupling versus Coulomb repulsion} 
\begin{abstract}
  We investigate via LDA+DMFT (local density approximation combined
  with dynamical mean field theory) the manifestation of correlation
  effects in a wide range of binding energies in the hole-doped family
  of Fe-pnictides $A$Fe$_2$As$_2$ ($A={\rm K}$, Rb, Cs) as well as the
  fictitious {\fr} and $a$-axis stretched {\cs}.  This choice of
  systems allows for a systematic analysis of the interplay of Hund's
  coupling $J_H$ and on-site Coulomb repulsion $U$ in multi-orbital
  Fe-pnictides under {\it negative} pressure.  With increasing ionic
  size of the alkali metal, we observe a non-trivial change in the
  iron $3d$ hoppings, an increase of orbitally-selective correlations
  and the presence of incoherent weight at high-binding energies that
  do not show the typical lower Hubbard-band behavior but rather
  characteristic features of a Hund's metal.  This is especially
  prominent in $a$-stretched {\cs}.
   We also find that the coherent/incoherent electronic behavior
of the systems is, apart from temperature, strongly dependent on $J_H$
and we provide estimates of the coherence scale $T^*$.
   We discuss these results in the framework of reported experimental observations.
\end{abstract}

\maketitle

\section{INTRODUCTION}
\label{sec:introduction}

The nature and degree of correlations in Fe-based superconductors has
been a subject of intensive discussions since the discovery of the
first high-$T_c$ iron pnictide superconductor in 2008~\cite{Hosono}.
A significant amount of work has concentrated on the description of
experimentally observed effects of correlation like
large effective masses or possible non-Fermi liquid
behavior~\cite{haulekotliar2009,Aichhorn2010,haulekotliar2011,medici2011,Yu2011,
ferber2012,ferber2012b}.  Due to the multi-orbital nature of Fe-based
superconductors, the Hund's coupling $J_H$ has been shown to play a
key role in determining the behavior of these
systems~\cite{haulekotliar2009,Aichhorn09,liebsch2011,medici2011,Yu2011,ferber2012,ferber2012b,georges2013}. However,
there is an ongoing debate regarding the role of $J_H$ versus the
on-site Coulomb repulsion $U$ and the interpretation of the correlated
nature of Fe-pnictides and
Fe-chalcogenides~\cite{haulekotliar2009,haulekotliar2011,medici2011,Yao2011,georges2013,Yu2013_5orb,medici2014,bascones2015}.
An important feature observed in many of these studies is that,
depending on the electronic filling, the Hund's coupling $J_H$ on the
one hand renders a moderately correlated system even more correlated
and pushes it into a bad metal regime, while on the other hand it can
also reestablish a metallic behavior, albeit orbital selective, in a
strongly correlated system~\cite{georges2013,bascones2015}.  The main
question for the Fe-based superconductors narrows down to which regime
of parameters do they belong to and how do correlations manifest in a
wide range of binding energies as a function of doping and/or
pressure?

Of special interest are the hole-doped $A$Fe$_2$As$_2$ ($A={\rm K}$, Rb,
Cs) end members of the 122 iron pnictide series. Compared to the
parent compound {\ba}, the substitution of Ba by K accounts for a
doping of one hole per formula unit and it is accompanied by a
complete suppression of any structural or magnetic phase
transition~\cite{paglione2010,boehmer2014} and by the appearance of
superconductivity at low temperatures. This behavior is
common~\cite{kihou2010,dong2010,eilers_thesis2014,zhang2015,hong2013}
to all hole-doped end members $A$Fe$_2$As$_2$ and they all seem to
have a nodal gap structure~\cite{hong2013,wang2013}, which is
different to the nodeless gap structure found in the parent {\ba}
system~\cite{ding2008,terashima2009,luo2009}.  Further, experimentally
Ba$_{1-x}$K$_x$Fe$_2$As$_2$ is thought to undergo a coherence-incoherence 
transition~\cite{werner2012,hardy2013,liu2014} as a function of
temperature, that has been interpreted in terms of a strong increase
in correlations~\cite{popovich2010,mu2009,terashima2010}.
Measurements of the Sommerfeld coefficient suggest that the hole-doped
end systems are among the strongest correlated 122 iron-pnictide
superconductors~\cite{hardy2013, eilers_thesis2014}, which is also
corroborated by theoretical investigations on
{\kf}~\cite{haulekotliar2011,hardy2013,backes2014,skorn_vollhardt2014,medici2014}.  The measured
Sommerfeld coefficient increases from {\ba} to {\kf} by more than an
order of magnitude~\cite{hardy2013,boehmer_thesis2014,storey2013} and
increases further as K is substituted by Rb and
Cs~\cite{shermadini2010,zhang2015}.  In view of these observations,
the hole doped end members $A$Fe$_2$As$_2$ provide an ideal background
for investigating correlation
effects as function of {\it negative} pressure.

In the present work we analyze via a combination of density functional
theory (DFT) in the local density approximation (LDA)
with dynamical mean field theory (LDA+DMFT) the electronic
structure of the series  $A$Fe$_2$As$_2$ ($A={\rm K}$, Rb, Cs)  
as well as the fictitious {\fr} and $a$-axis stretched {\cs} in
an extended range of binding energies. 
The combination of density functional theory with many body methods
such as dynamical mean field theory has proven to be a very helpful
tool to investigate electronic correlation 
effects~\cite{dmft:anisimov1997,dmft:held2006,dmft:kotliar2006,dmft:lechermann2006}.
  While a compression of the unit
cell usually decreases correlation effects in the 122 iron
pnictides~\cite{Diehl2014,Mandal2014,Guterding2015}, similarly, the opposite is
to be expected when lattice parameters are expanded since the larger
interatomic distances should reduce the hybridization of neighboring
atomic orbitals and lead to stronger localization of the electronic
states.  We show in this study that these considerations are correct
only at first sight; actually, the strong correlation effects in these
systems are mostly governed by a subtle interplay of $J_H$ and $U$.

\section{RESULTS and DISCUSSION}
\label{sec:results}

We performed DFT(LDA) and LDA+DMFT 
calculations on experimental structures
of {\kf}, {\rb} and {\cs}~\cite{struct_ref} in tetragonal symmetry.  Additionally, we
prepared a fictitious structure for {\fr} with lattice parameters
and As height obtained by extrapolating the experimental
parameters for AFe$_2$As$_2$ (A = K, Rb, Cs) (see Appendix A). 
Furthermore, we considered an $a$-axis stretched {\cs} structure by
 slightly expanding the lattice of {\cs} in the $a$-$b$ plane
to mimic an experimental biaxial stretching of the crystal. This may 
be achieved experimentally by epitaxial growth of {\cs} on a substrate
with slightly larger lattice parameter. The structural trends of this series are an
expansion along $a$ and $c$ and a decrease of the As height. A detailed explanation of the
structures and computational methods is given in Appendix \ref{appendix:methods}.

\subsection {Momentum-resolved spectral function $A({\bf k},\omega)$}
\label{sec:kdep_spec}

\begin{figure}[t]
\includegraphics[width=1\linewidth]{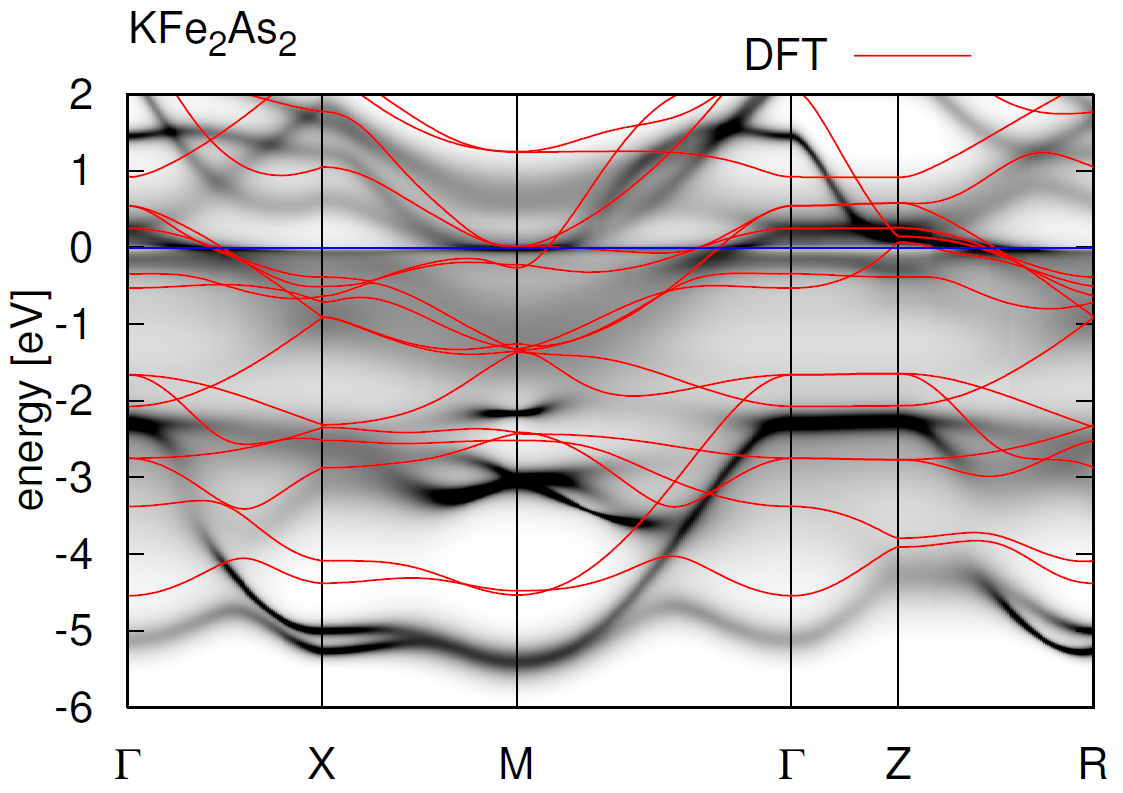}
\includegraphics[width=1\linewidth]{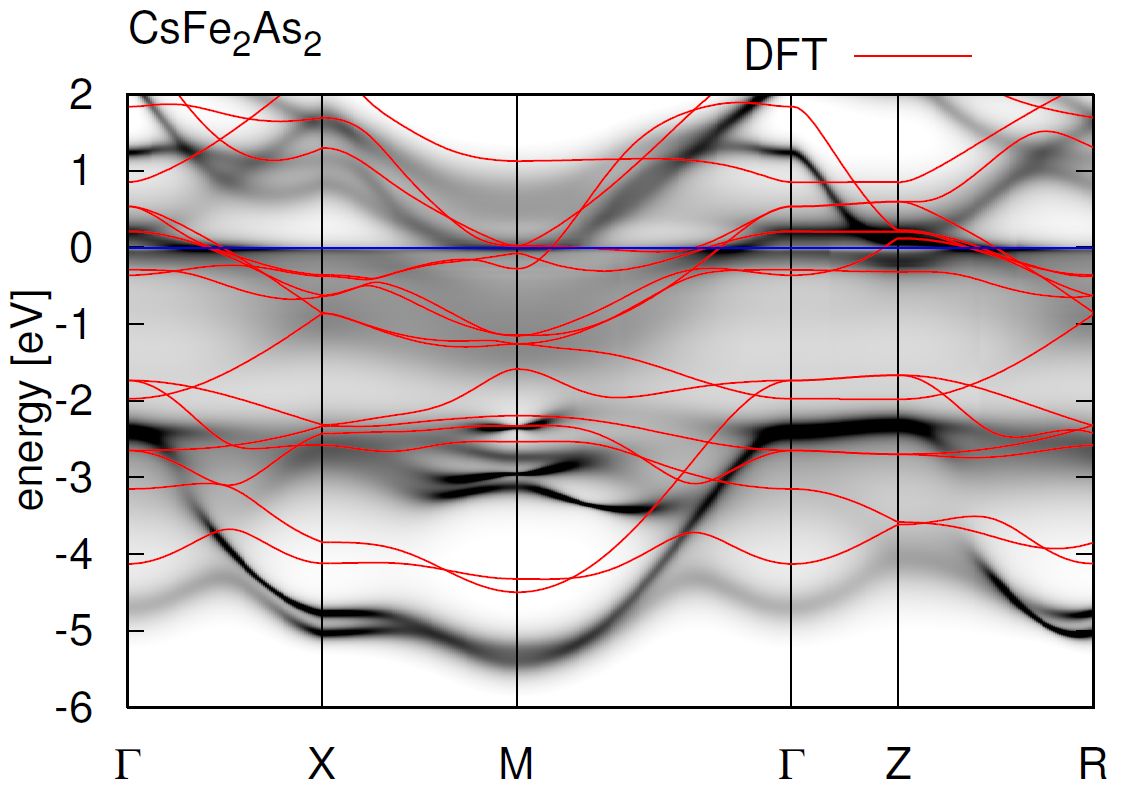}
\caption{(Color online) Momentum-resolved spectral function of {\kf} (top) and
  {\cs} (bottom). Strong correlations in these materials introduce
  renormalization effects as well as broadening in the spectral
  function due to finite quasiparticle lifetimes compared to DFT(LDA). We
  observe a notable broadening and suppression of spectral features
  especially in the energy range $[-2.0,-0.5]$~eV.}
\label{fig:bs_k_cs_large}
\end{figure}

In Fig.~\ref{fig:bs_k_cs_large} we show the momentum-resolved spectral
function for {\kf} and {\cs} as obtained within LDA+DMFT (gray density
plot) at a temperature $T=145$~K and compare it to the DFT bandstructure
(red).  We estimate for both systems Fe $3d$ effective masses
$m^*/m_{\rm LDA}$ between 2.2 and 4.1 (depending on the orbital) which lead to a
renormalization of the DFT(LDA) band energies and overall reduction of
bandwidth.  The average effective mass increases slightly from $2.89$
to $2.95$, which indicates an increase in correlation along the
$A$Fe$_2$As$_2$ series due to enhanced localization of electrons on
Fe.  Already at the Fermi level we obtain diffuse structures
corresponding to incoherent quasiparticle excitations with finite
lifetimes.  At energies below $-0.5$~eV all coherent features are
basically washed out due to correlation.  This effect is present in
all systems in the series {\kf}$\rightarrow${\cs}, where we observe an
energy range of strong broadening and depletion of spectral weight
compared to the DFT bandstructure between $[-2.0,-0.5]$~eV.  At
energies below $-2$~eV coherent features become visible again, which
correspond to the As $p$ states that partially hybridize with Fe $3d$
states. Even though the self-energy in DMFT has no momentum dependence,
an effective momentum dependence is present in the
results due to the momentum dependent orbital
character of the original DFT bands. This  leads to k-dependent
 broadening
effects in LDA+DMFT.

\subsection{ Density of states at the Fermi level and orbital-resolved
electronic filling}
\label{sec:dos_ef}

\begin{figure}[t]
\includegraphics[width=0.85\linewidth]{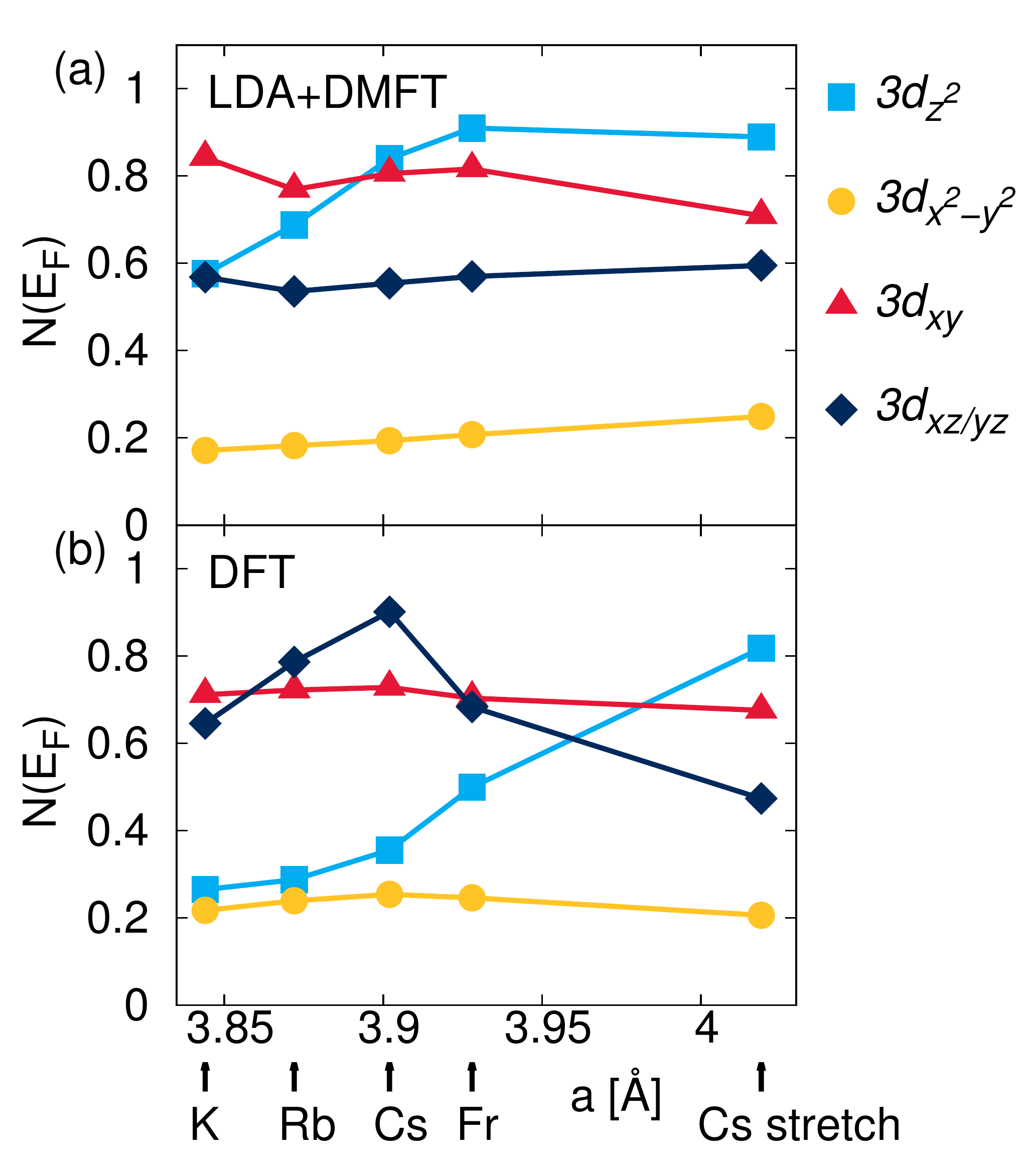}
\caption{(Color online) The density of states of the Fe $3d$ orbitals
  at the Fermi level as obtained from (a) LDA+DMFT and as comparison
  with (b) DFT. The electronic correlations induce a marked deviation
  from the DFT results: The contribution of the Fe $3d_{z^2}$ orbital
  is strongly enhanced, while a more pronounced decrease of the Fe
  $3d_{xy}$ orbital is found towards the end system of stretched
  {\cs}. Also, the trend in the Fe $3d_{xz/yz}$ orbital in DFT is
  completely evened out in LDA+DMFT.}
\label{fig:dos_ef_comp}
\end{figure}

We first analyze the manifestation of correlation effects near the
Fermi level.  For that we compare in Fig.~\ref{fig:dos_ef_comp}
LDA+DMFT (Fig.~\ref{fig:dos_ef_comp} (a)) with DFT(LDA)
(Fig.~\ref{fig:dos_ef_comp} (b)) orbital-resolved density of states at
the Fermi level $N(E_F)$ for all studied systems. The LDA+DMFT
calculations show an increasing and pronounced dominance of $3d_{z^2}$ contribution
at $E_F$ along the series.  This is in contrast to the DFT results
where the Fe $3d_{z^2}$ orbital contribution also increases, but is much lower and only 
significant for $a$-axis stretched {\cs} with shortest As height.  The nonmonotonous
behaviour of the Fe $3d_{xz/yz}$ contribution in DFT is a result of
the special electronic structure of these systems at the M point. This
feature is greatly suppressed by broadening effects in the LDA+DMFT
calculation.

\begin{figure}[t]
\includegraphics[width=0.85\linewidth]{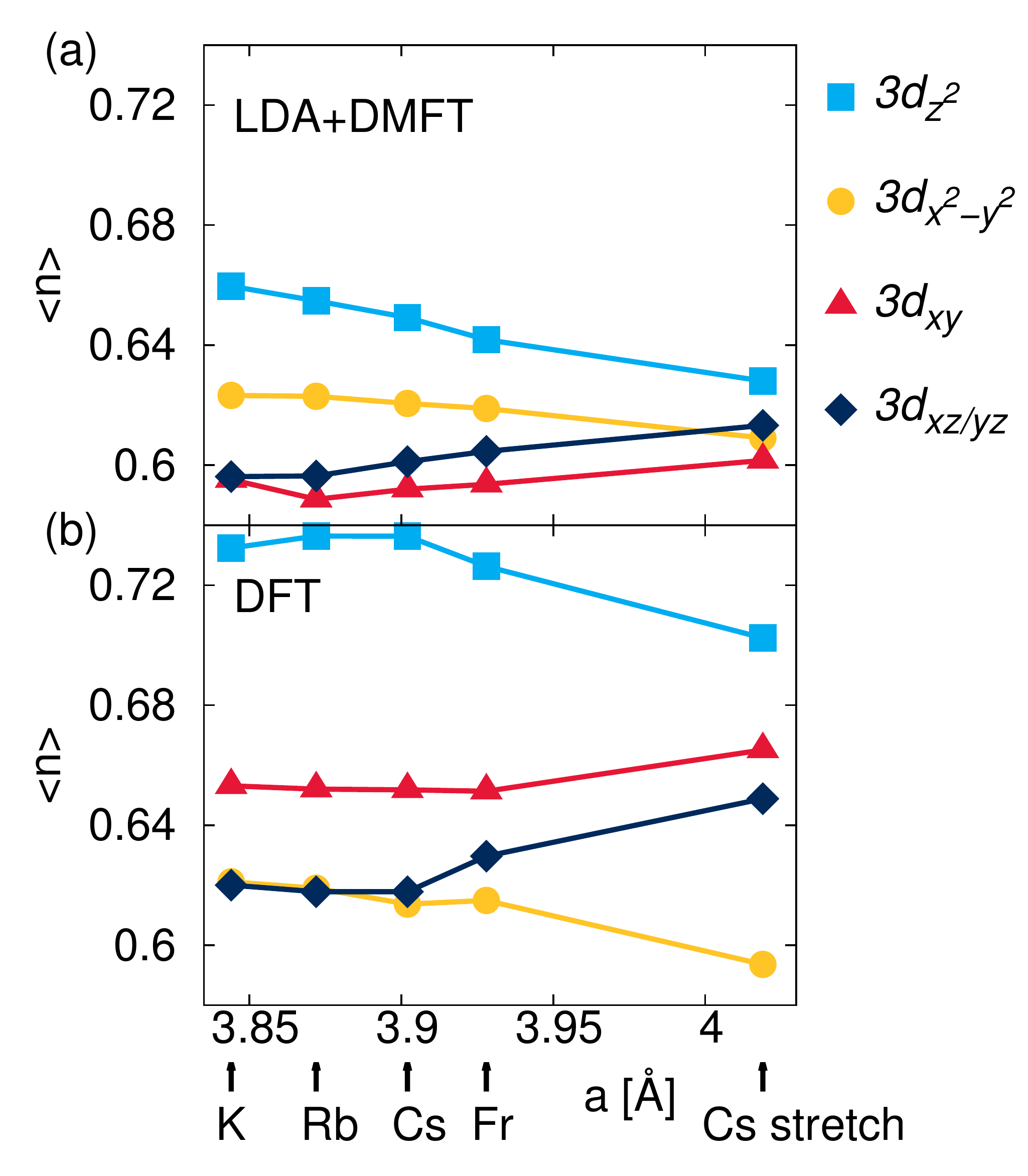}
\caption{(Color online) Filling of the Fe $3d$ orbitals as obtained
  from (a) LDA+DMFT and (b) DFT(LDA). 
Correlations  reduce the overall Fe $3d$ filling,
  with the most correlated $3d_{xy}$ and $3d_{xz/yz}$ orbitals
 being the closest 
  to half filling in contrast to the DFT result.
  Along the alkali metal series, the occupation of Fe
  $3d_{z^2}$ and $3d_{x^2-y^2}$ orbitals reduces, while that of the
  $3d_{xy}$ and $3d_{xz/yz}$ orbitals increases after a local minimum
  in {\rb}, which has the smallest $3d_{xy}$ occupation of all systems
  studied.}
\label{fig:filling_comp}
\end{figure}

Linked to these results is the behavior of the LDA+DMFT
orbitally-resolved electronic filling (see
Fig.~\ref{fig:filling_comp}). We observe that
inclusion of correlation effects not only reduces the overall
Fe $3d$ filling due to the additional cost of the Coulomb interaction energy
for doubly occupying a certain orbital, but also
introduces orbital-dependent effects
(see section \ref{sec:meff_lifetime}).
While the Fe $3d_{z^2}$ and $3d_{x^2-y^2}$ filling decreases along the series,
 it increases for
$3d_{xz/yz}$ and $3d_{xy}$.

\subsection{Spectral function $A(\omega)$}
\label{sec:spec_function}

\begin{figure}[t]
\includegraphics[width=1\linewidth]{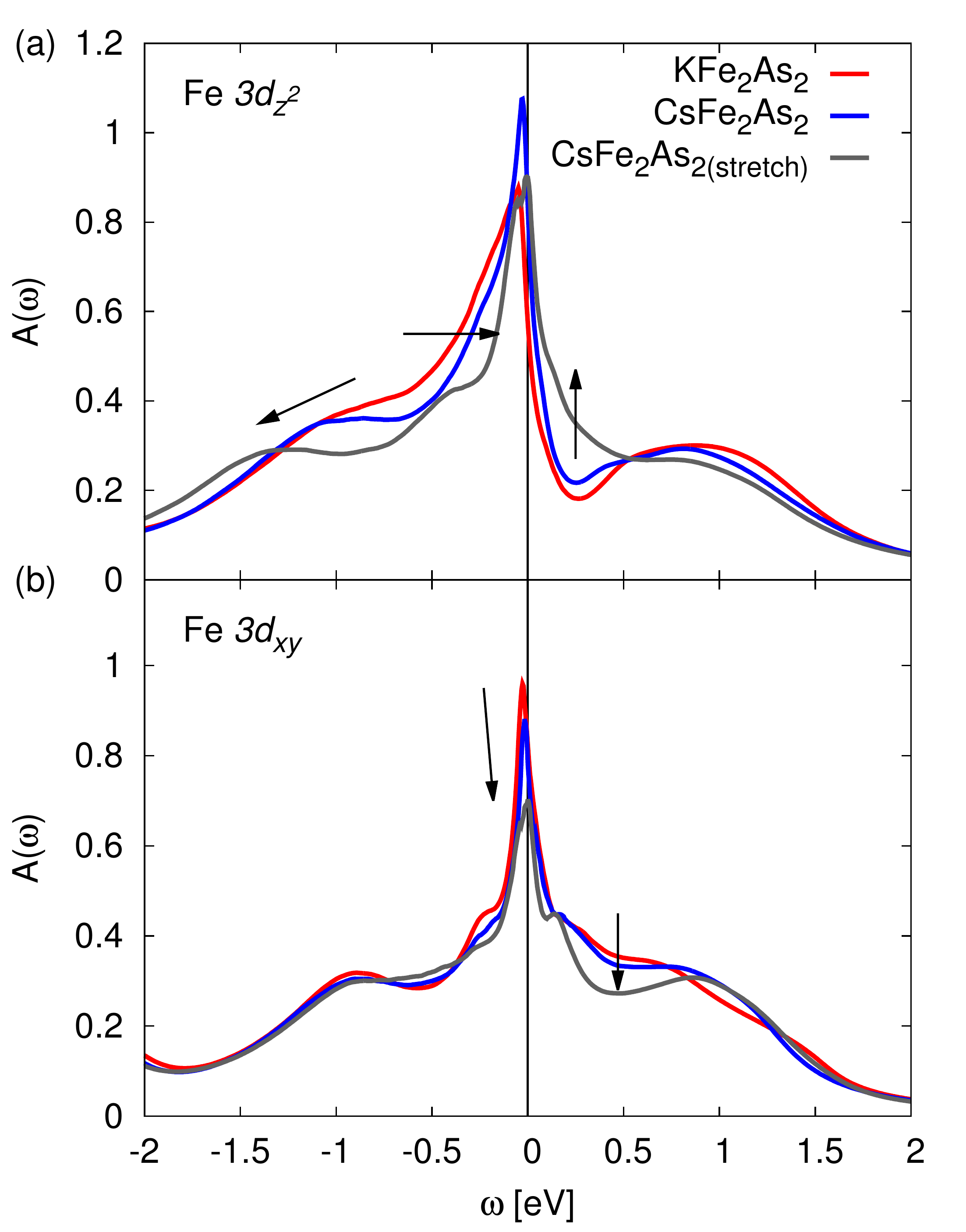}
\caption{(Color online) The density of states for (a) the Fe $3d_{z^2}$
  orbital and (b) the Fe $3d_{xy}$ orbital as obtained from LDA+DMFT
  for the three compounds {\kf}, {\cs} and an $a$-axis stretched 
  {\cs}.  Note the emergence of a Hubbard-like peaks around $-1.2$~eV
  and $+1$~eV for $3d_{z^2}$ and around $-1$~eV and $+1$~eV for
  $3d_{xy}$. With arrows we mark the trend of the changes in the
  spectral function along the series.}
\label{fig:z2_doslarge}
\end{figure}

In order to understand the origin of the changes in $N(E_F)$ and
orbital filling, we show in Fig.~\ref{fig:z2_doslarge} the local
density of states (spectral function $A(\omega)$) for Fe $3d_{z^2}$
and Fe $3d_{xy}$ in the energy range $[-2, 2]$~eV for the
representatives {\kf}, {\cs} and $a$-axis stretched {\cs}.  We chose
these two orbitals since they are the most affected along the series:
$3d_{z^2}$ because of the decrease in As height from {\kf} to {\cs}
and $3d_{xy}$ because of the increase in the orbital localization with
increasing $a$ lattice parameter.  Fe $3d_{z^2}$ shows a shift of
orbital weight to negative energies (high binding energies) and a
narrowing of the quasiparticle-like peak structure at $E_F$ from {\kf}
to {\cs}.  Since the electronic filling of the iron $3d$ orbitals is
larger than half-filling in these systems, the quasiparticle-like peak
is located close to but below the Fermi level.  In Fe $3d_{z^2}$ due
to the reduction of the filling from 0.66 in {\kf} to 0.63 in
stretched {\cs} (see Fig.~\ref{fig:filling_comp}) caused by an
increase of electronic correlations, the quasiparticle peak-like
structure is shifted even closer towards the Fermi level along the
series, which in turn leads to the observed increase of the density of
states at the Fermi level (Fig.~\ref{fig:dos_ef_comp}(a)).  Such a
shift of the quasiparticle peak-like structure in {\kf} was already
noted in Ref.~\onlinecite{skorn_vollhardt2014}. For the $3d_{xy}$
orbital, the quasiparticle peak is much closer to the Fermi level
since the filling is closer to half filling at $n_{xy}\approx 0.59$.
This orbital shows a strong suppression of the quasiparticle peak (up
to $\approx 30\%$) from {\kf} to the $a$-axis stretched {\cs} which
points to an important increase of decoherence along the series. This
increase of decoherence will be studied in more detail in the next
section.  Reduction of the maximum of the quasiparticle peak combined
with a slight change of its position results in the almost constant
density of states at the Fermi level for the $3d_{xy}$ orbital
observed in Fig.~\ref{fig:dos_ef_comp} (a).

Additionally, a shoulder-like feature appears in the spectral function
at $1$~eV and $-1$~eV in {\cs},  closely
resembling the typical spectral function shape of a quasiparticle peak and a lower and
upper Hubbard band.  These features do not correspond to any property
found in the non-interacting DOS and are purely an effect of
correlations and, at first sight, are similar to the emergence of
Hubbard bands as a function of $U$ in strongly correlated
systems. This is also in agreement with the $3d_{xy}$ orbital being
the strongest correlated one, whereas these features are far less
developed in the $3d_{z^2}$ and other orbitals (see Appendix
\ref{appendix:dos_other_orbs}).  However, as we will show in section
\ref{sec:U_J_dependence}, these peaks do not behave as expected for 
Hubbard bands in a one-band Hubbard model.

\subsection{ Effective masses and quasiparticle lifetimes}
\label{sec:meff_lifetime}

\begin{figure}[t]
\includegraphics[width=1\linewidth]{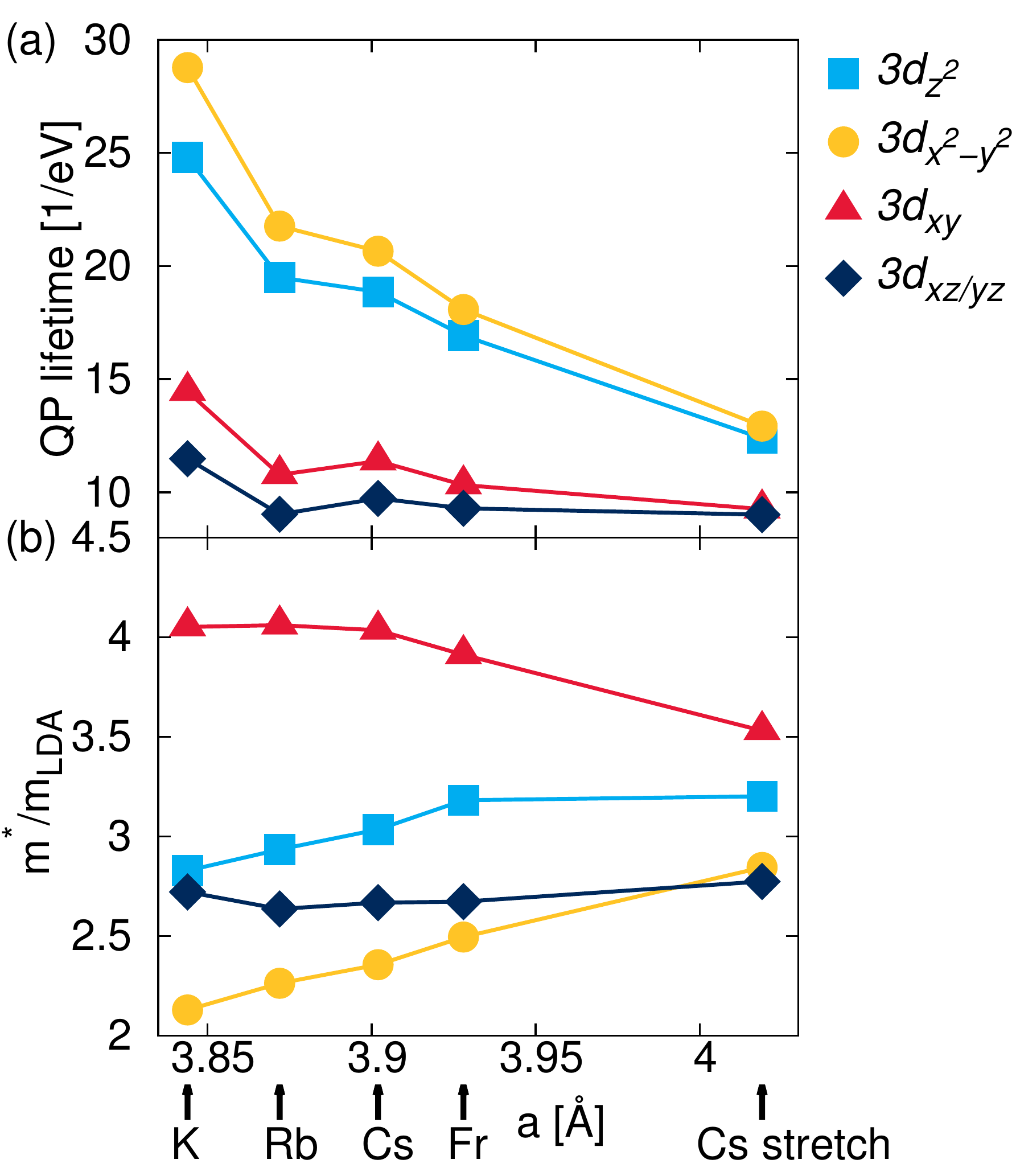}
\caption{(Color online) (a)  Quasiparticle lifetimes  given by
  $\tau_m= -(Z_m \mathrm{Im} \Sigma_m(i0^+))^{-1}$
of the Fe $3d$ orbitals along the $A$Fe$_2$As$_2$ series and (b) the corresponding 
effective masses $\frac{m^*}{m_{\rm LDA}}$.}
\label{fig:meff_life}
\end{figure}

In order to quantify the change in correlation along the series
$A$Fe$_2$As$_2$ we plot in Fig.~\ref{fig:meff_life} the orbitally
resolved quasiparticle lifetimes and effective masses
$\frac{m^*}{m_{\rm LDA}}$.  The effective masses for Fe $3d_{z^2}$ and
$3d_{x^2-y^2}$ increase along the $A$Fe$_2$As$_2$ series but remain
constant or even slightly decrease for $3d_{xz/yz}$ and $3d_{xy}$.
This last result cannot be explained solely by the behavior
of the DFT-derived 
 tight binding parameters (see
Appendix~\ref{appendix:10band_tb} and \ref{appendix:16band_tb}).

Information on the origin of this behavior can be obtained from the
quasiparticle lifetimes which strongly decrease for all orbitals along
the series (Fig.~\ref{fig:meff_life} (a)).  This indicates that the
coherent quasiparticle picture, being the basis for the calculation of
the effective masses, becomes less appropriate along the
$A$Fe$_2$As$_2$ series, and accordingly, the effective masses obtained
by this procedure are an underestimation of the true value. This
result shows that already at the temperature of $T=145$~K these
systems are quite incoherent. With increasing lattice parameter
incoherence significantly increases, albeit leading to an orbital
dependent change in localization of the Fe $3d$ electrons.  In
particular, we find a pronounced decrease of the quasiparticle
lifetime from {\kf} to {\rb}, which we attribute to the competing
effects of decrease of Fe-Fe direct hopping and increase in Fe-Fe
indirect hopping through As, which seem to have a crossover point
between {\rb} and {\cs} (see Appendix \ref{appendix:10band_tb}).

\begin{figure}[t]
\includegraphics[width=1.0\linewidth]{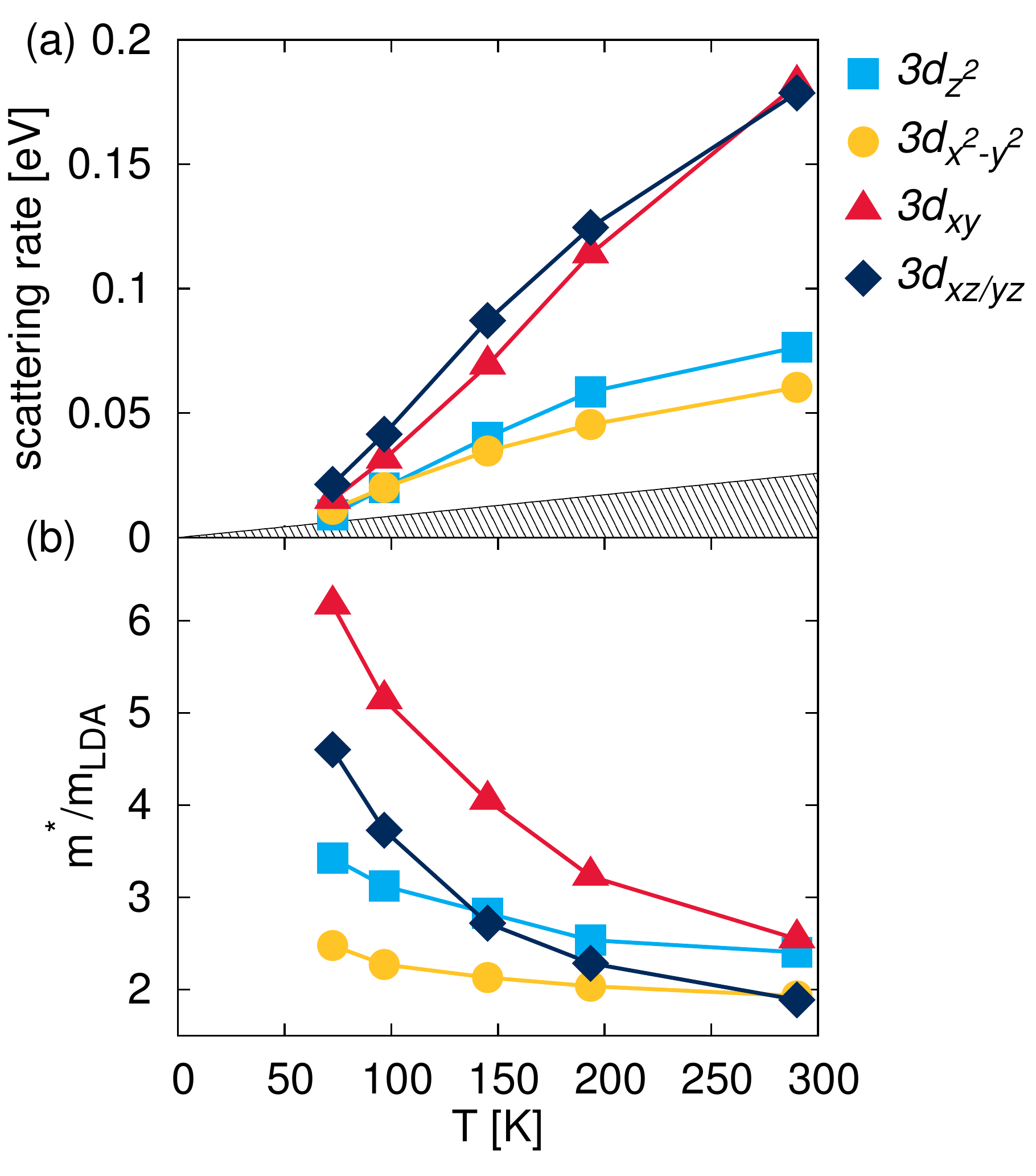}
\caption{(Color online) Temperature dependence of the
(a) scattering rates $-Z_m \mathrm{Im} \Sigma_m(i0^+)$  of the Fe $3d$ orbitals of {\kf}.
The shaded area indicates the coherent domain.
The coherence temperature estimate $T^*$ as deduced from the
  scattering rate is quite low, located around $50$~K.
 (b) The corresponding effective masses.
  }
\label{fig:tempDep}
\end{figure}

In Fig.~\ref{fig:tempDep} we show the temperature dependence of the
quasiparticle scattering rates and effective masses of {\kf}. 
We find a similar temperature dependence as also observed for
multi-orbital SrRuO$_3$ and CaRuO$_3$ in Ref.~\onlinecite{Millis2015}. As the
temperature is lowered in the calculation, the quasiparticle picture,
which is suppressed at high temperatures, is partially restored,
leading to an exponential increase in the quasiparticle lifetimes.
Still, even at the lowest studied temperature of $T=72$~K the width
(scattering rate) of the quasiparticle peak $-Z_m \mathrm{Im}
\Sigma_m(i0^+)$ is still larger than the temperature, indicating that
coherent quasiparticle excitations are still in the minority.  From
our results for the scattering rate we estimate that the temperature
$T^*$ for the incoherent-coherent phase transition for our chosen
interaction parameters $U=4$~eV, $J_H=0.8$~eV is around $50$~K, and is
even lower for {\rb} and {\cs}.  This is in qualitative agreement with
previous LDA+DMFT estimates for {\kf}~\cite{skorn_vollhardt2014} and
with magnetic susceptibility and thermal expansion measurements which
predict the coherence scale of {\kf} to be around
$100$~K~\cite{liu2014,hardy2013,boehmer_thesis2014} and even lower for
{\rb} and {\cs}~\cite{eilers_thesis2014,boehmer_thesis2014}.  We
expect that inclusion of the full rotationally invariant Hund's
coupling, which is beyond the scope of the present work, will shift the
transition to higher temperatures in the
calculation~\cite{biermann_rotinv,haulekotliar2009,Aichhorn2010} and,
therefore, closer to experiment.  The consequences of inclusion
of rotationally invariant interactions have been extensively discussed
in previous works in the framework of the two-band~\cite{Pruschke}
and three-band~\cite{Antipov} Hubbard models. The computational effort
of including these terms in the five-band cases studied here 
is beyond the scope of the present work. However, we performed some benchmarking
calculations with fully rotational Hund's coupling
 and find that the coherence transition
shifts to higher temperatures as found in other studies~\cite{biermann_rotinv,haulekotliar2009,Aichhorn2010}. In this sense the presented results
can be considered as a lower bound to the experimental observations.
 
 The increase of effective
masses at lower temperatures in our calculation is precisely the
effect of restoring the coherent quasiparticle picture, so that the
effective masses at the lowest temperature ($72$~K) we investigated
can be considered as the closest approximation to the true
values, i.e. $6.1$ ($3d_{xy}$), $4.6$ ($3d_{xz/yz}$), $3.4$
($3d_{z^2}$) and $2.5$ ($3d_{x^2-y^2}$).

Combining these observations along the series we conclude that  alkali
122 systems show typical signs of  strong correlations but, in the studied range of temperatures,
 are actually quite
deep in the incoherent bad metal region  with a well defined, albeit
strongly suppressed, quasiparticle peak.

\subsection{Nature of the Fe $3d$ wavefunction}
\label{sec:sectorstatistics}

\begin{figure}[t]
\includegraphics[width=0.65\linewidth]{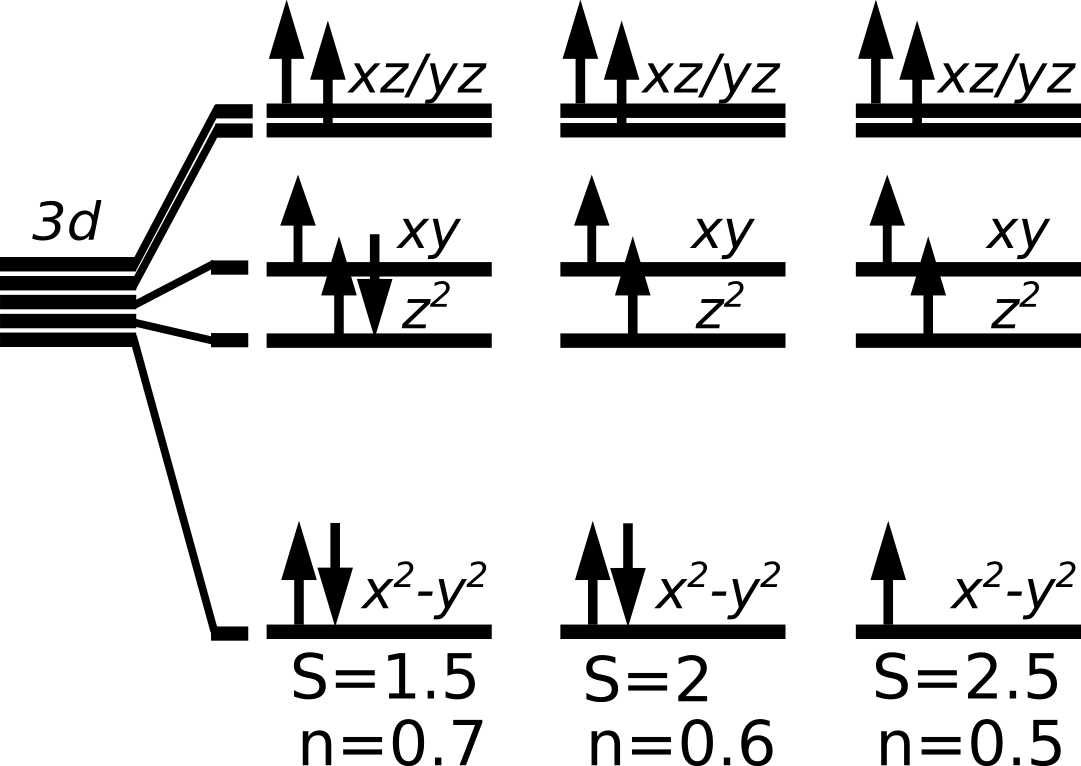}
\caption{Sketch of some of the most likely Fe atomic orbital configurations in hole-doped
122 iron pnictide systems. The electronic filling is indicated by $n$ and the total spin by $S$.}
\label{fig:orbStates}
\end{figure}

We proceed now with an analysis of the wave function in terms of the
Fe atomic basis states, similar to what has been done for other
pnictide systems~\cite{haulekotliar2009,haulekotliar2011}.  In
Fig.~\ref{fig:orbStates} we show a sketch for a few typical orbital
configurations in these systems with the orbital splittings obtained from 
the downfolded charge self-consistent LDA Hamiltonian.
Since the nominal electronic filling
for Fe $3d$ for these systems is 5.5 electrons per Fe, one
would expect atomic Fe $3d$
states with 5 and 6 electrons to be the most likely states. 
However, as shown in Fig.~\ref{fig:filling_comp}, the actual
average Fe $3d$
filling is slightly larger than 6 and this can be analyzed in the
histogram of Fig.~\ref{fig:orbStates} . For the
interacting system there is a non-trivial competition between the
energy contribution due to the crystal field splitting, which prefers
to occupy the lowest states first, the on-site $U$ interaction, which
tends to decrease the filling of the localized states and the Hund's
coupling $J_H$, which prefers orbital states with maximum total spin.
When the Hund's coupling $J_H$ is large compared to the total crystal
field splitting, the high-spin states will have the highest
probability and the low-spin states will be suppressed. This is indeed
true for the hole-doped 122 iron-pnictides.

\begin{figure}[t]
\includegraphics[width=1\linewidth]{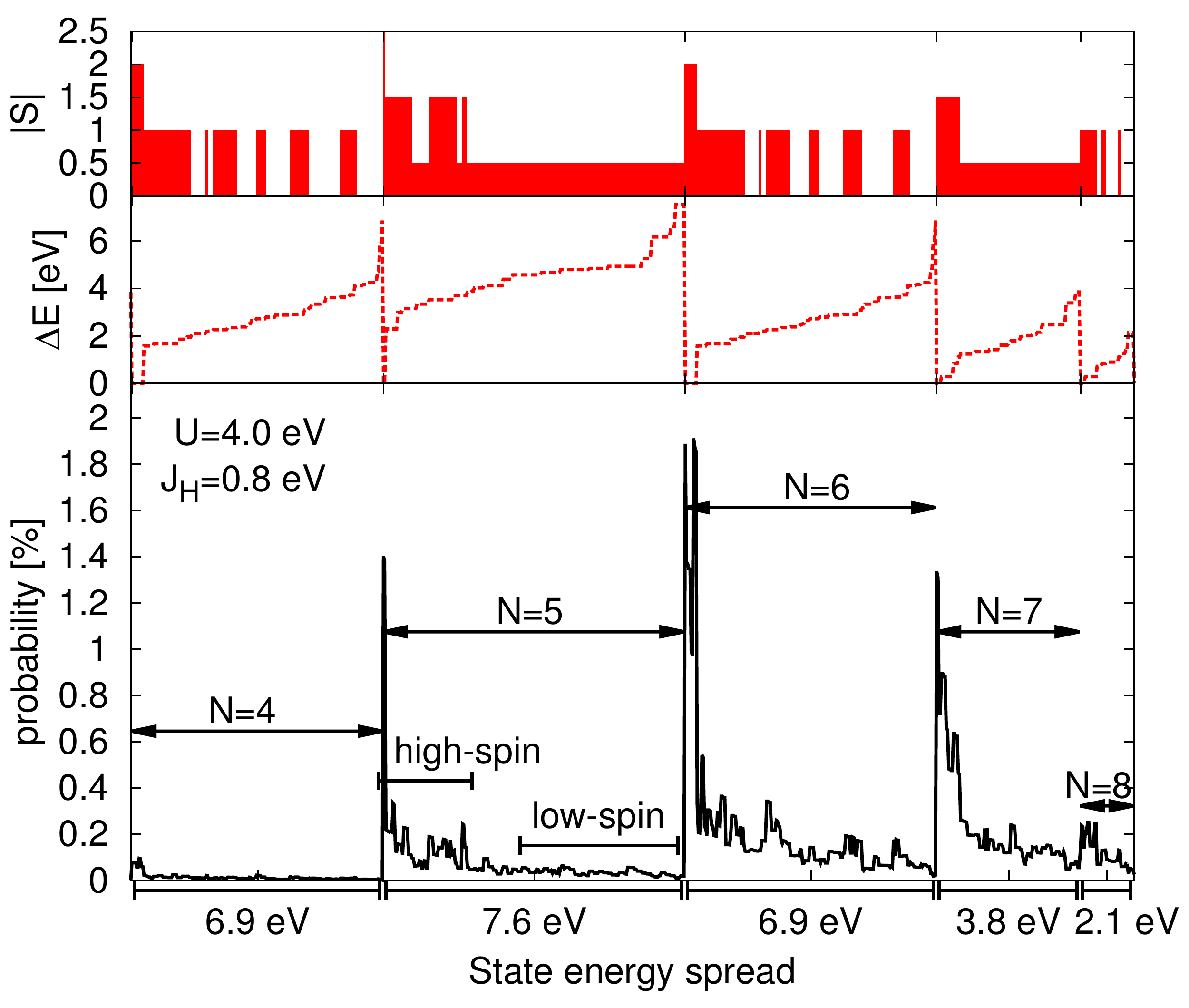}
\caption{(Color online) The histogram of the Fe $3d$ atomic state for
  {\kf} at $T=145$~K. The probability corresponds to the fraction of
  time the Fe $3d$ orbitals spend in one of the $2^{10}=1024$ possible
  states. Within the interval of constant electron number $N$ the
  states are sorted by increasing energy, i.e. the leftmost states
  within an interval correspond to high-spin states, while the
  rightmost states correspond to low-spin states. }
\label{fig:state_histogram}
\end{figure}

\begin{figure}[t]
\includegraphics[width=0.9\linewidth]{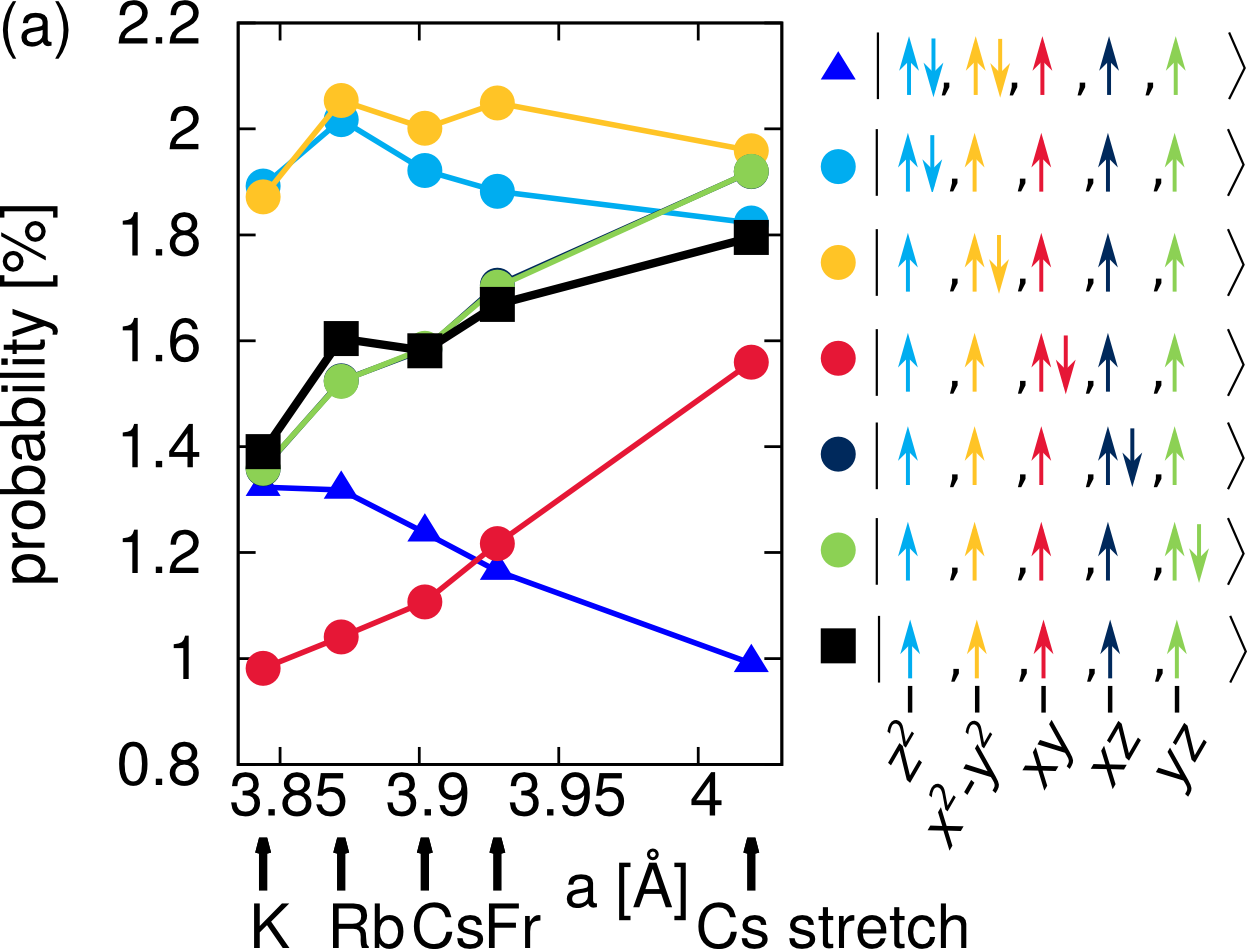}
\includegraphics[width=0.9\linewidth]{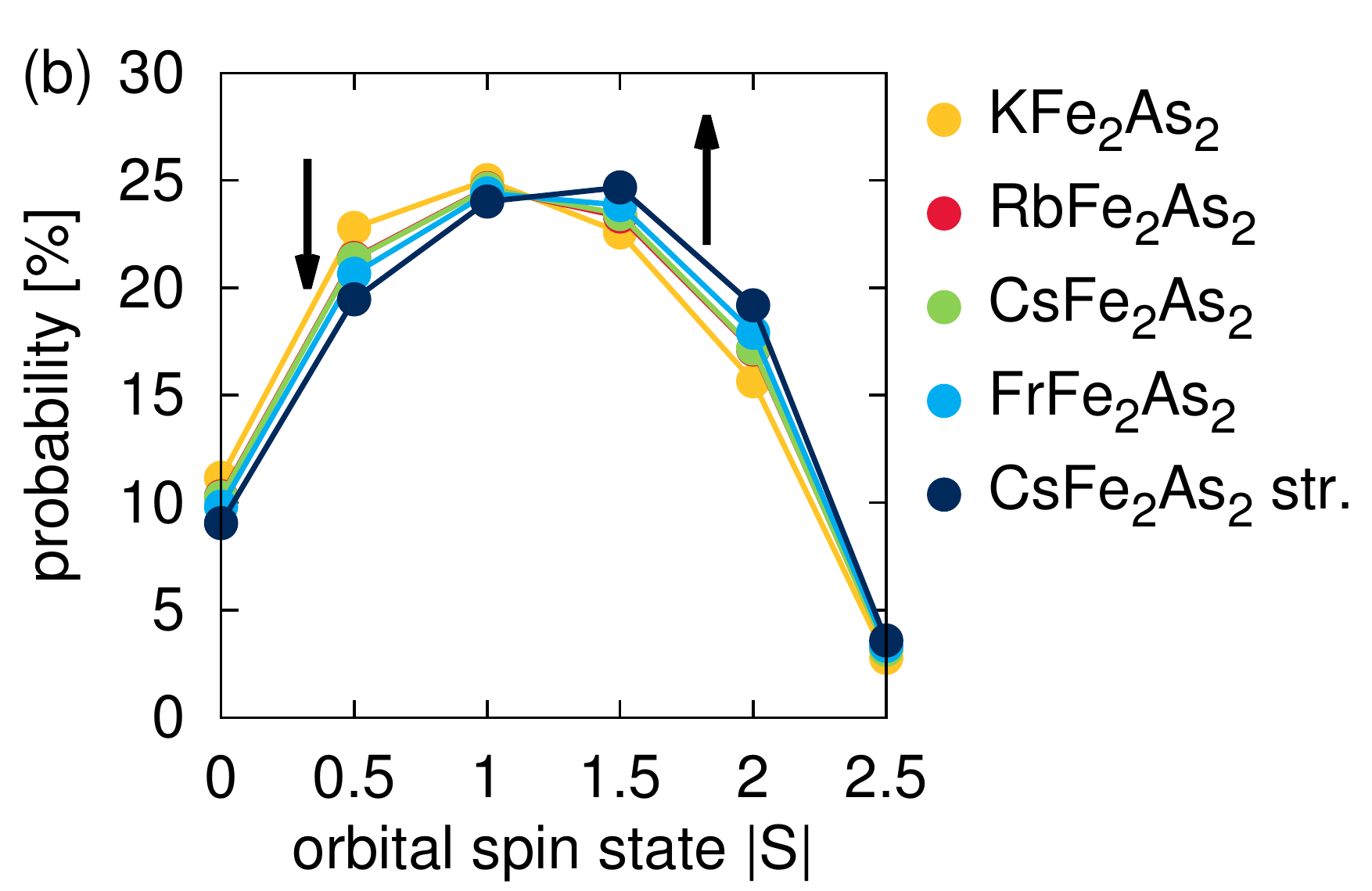}
\caption{(Color online) (a) Probabilities of the most
  likely atomic states of the Fe $3d$ atomic orbitals for the $A$Fe$_2$As$_2$ series.  The
  probability corresponds to the fraction of time the atom in the DMFT calculation spends in a
  specific state.  (b) The summed probability of all atomic
  states with a given total spin $S$. High-spin states become more
  likely for increasing lattice parameter $a$, while the probability
  of low-spin states is reduced. }
\label{fig:sectorStats}
\end{figure}

In Fig.~\ref{fig:state_histogram} we show for {\kf} the atomic histogram of the
Fe $3d$ shell, i.e. the projection of the wave function onto the Fe
$3d$ atomic basis states. The states are sorted by the number of
electrons $N$, and inside the interval of constant filling the states
are sorted by energy. Because of the Hund's coupling, the leftmost
states in such an interval correspond to the high-spin states,
while the rightmost states correspond to low-spin states.  The
probability assigned to each state corresponds to the fraction of time
in the calculation the $3d$ orbitals spend in a specific configuration. Due to the
Hund's coupling the high-spin states clearly dominate the
histogram, although their probabilities are quite low with less than
$2\%$. Even for the low-spin states at higher energy the probability, while  
being smaller, is never close to zero, which causes the Fe $3d$ orbitals
to visit a large number of accessible states over
time even when they are much higher in energy.  Since all these states
 contribute with a finite fraction to the
Fe $3d$ spectral function, no well defined atomic like excitations can
be expected. This leads to the observed suppression of the
quasiparticle peak and  subtle high-binding energy features in the
spectral function without well defined Hubbard bands, which is the
distinctive property of Hund's metals~\cite{haulekotliar2011,bascones2015}.

This behavior becomes even more pronounced when we perform this analysis
along the $A$Fe$_2$As$_2$ series. 
In Fig.~\ref{fig:sectorStats}~(a) we analyze  the most likely atomic
states from the histogram of the Fe $3d$ orbitals for $A$Fe$_2$As$_2$.
 For all systems, six
out of the seven most likely atomic states are solely composed of the
maximal high-spin states with $S=2.5$ and $S=2$.  For the earth alkali
(undoped) 122 iron-pnictides like BaFe$_2$As$_2$ the atomic ground
state of the Fe atom with a valence charge of $6$ has a maximum
possible spin of $S=2$. Since for the systems studied here one
electron per formula unit has been removed by hole doping, the probability for
the fully polarized half-filled $S=2.5$ state with 5 electrons in the
Fe $3d$ orbitals is among the most likely states with a comparably
high probability of $1.4\%$ (or $2.8\%$ when accounting for spin
degeneracy) in {\kf}, that increases up to $1.8\%$ (or $3.6\%$) in the
stretched {\cs} system. Along the series  the low-spin states become
suppressed while the high-spin states increase in probability, as can
be seen in Fig.~\ref{fig:sectorStats}~(b). 

Generally, the systems can reduce their energy by assigning a higher
probability to high-spin states due to the Hund's coupling $J_H$.
This leads to a significant increase of localization caused by the
orbital blocking mechanism~\cite{haulekotliar2011}.  Since $J_H$
enforces a high-spin state, orbital mixing is greatly suppressed
compared to a vanishing Hund's coupling where high- and low-spin
states would have equal energies and, therefore, probabilities. This
is the typical behavior of a so-called Hund's metal, in which the
electronic correlations are much more sensitive to the value of $J_H$
than to the on-site Coulomb term $U$.  Therefore, in the hole-doped
end systems like {\cs} and especially the $a$-axis stretched {\cs} the
Hund's coupling becomes the most important interaction that governs
the physical properties of these systems.


\subsection{Dependence on $U$ and $J_H$}
\label{sec:U_J_dependence}

\begin{figure*}[t]
\includegraphics[width=1\linewidth]{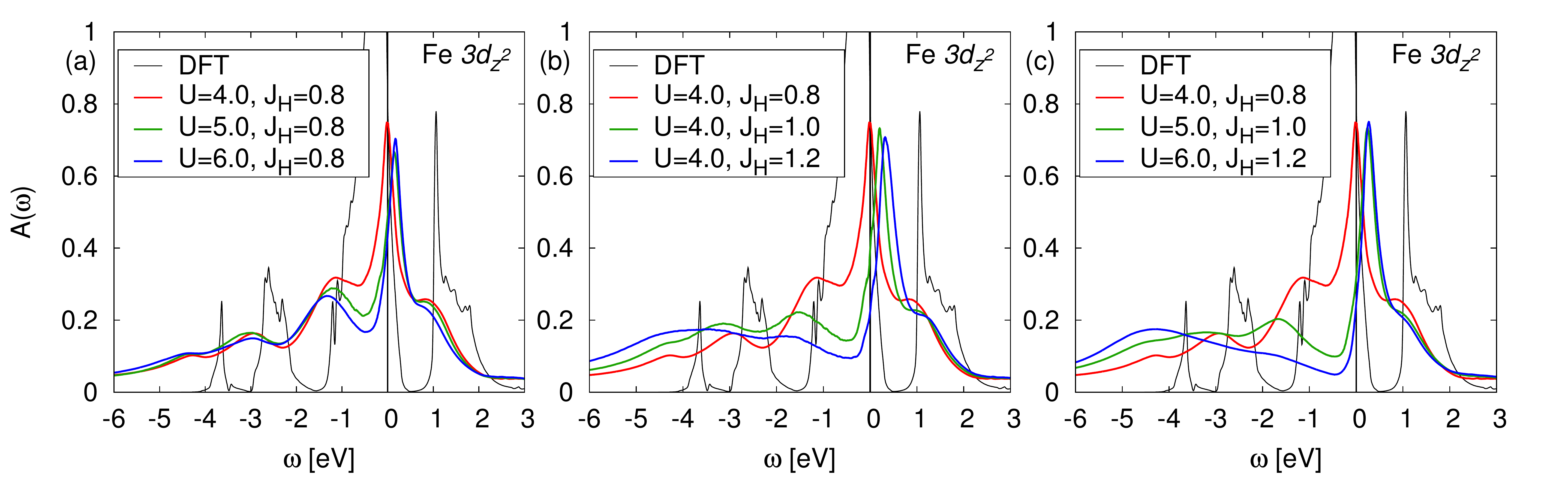}
\caption{(Color online) The density of states for the Fe $3d_{z^2}$
  orbital of the stretched {\cs} compound as a function of the on-site
  Coulomb repulsion $U$ and Hund's coupling $J_H$. (a) An increase
  only in $U$ leads to an increase in renormalization, i.e. effective
  masses and a pronounced Hubbard-like peak at $-1.5$~eV but no other
  qualitative changes are observed.  (b) The Hund's coupling $J_H$
  greatly increases the decoherence of the electronic states at the
  Fermi level and leads to a significant shift of spectral weight down
  to lower energies. (c) The combined effect of $U$ and $J_H$ is
  qualitatively very similar to an increase in $J_H$ alone.}
\label{fig:z2_UJ_dep}
\end{figure*}

\begin{figure*}[t]
\includegraphics[width=1\linewidth]{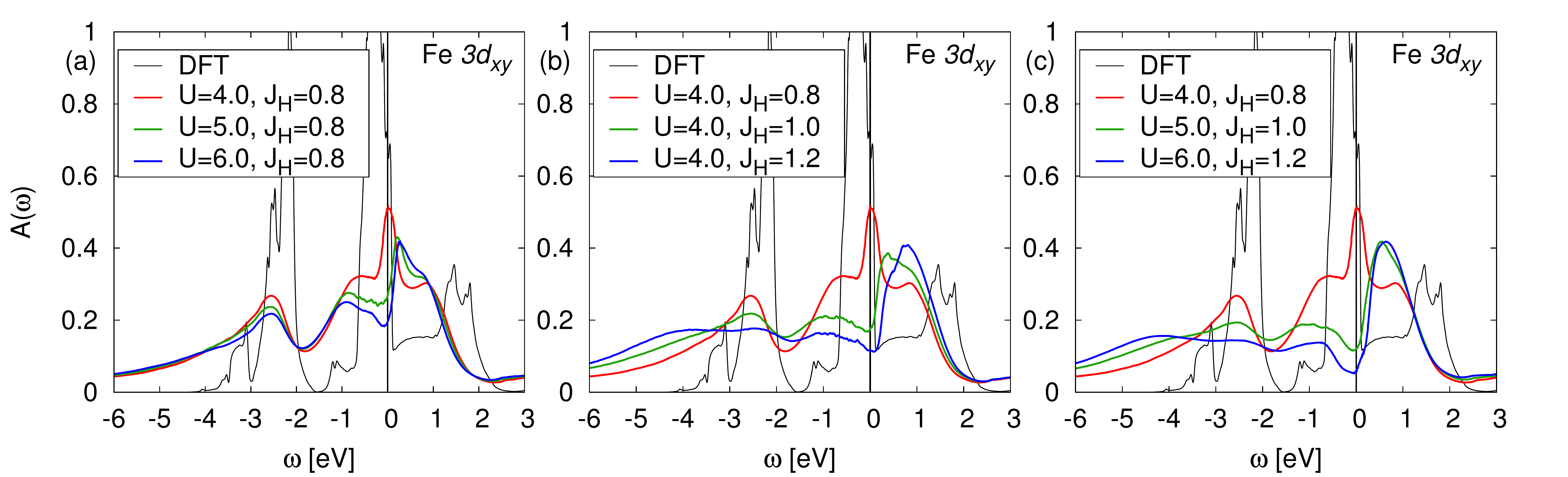}
\caption{(Color online) The density of states for the Fe $3d_{xy}$
  orbital of the stretched {\cs} compound as a function of the on-site
  Coulomb repulsion $U$ and Hund's coupling $J_H$. Similar to the Fe
  $3d_{z^2}$ orbital, we see that the shift of
  spectral weight to lower energies is almost exclusively dependent
  on the Hund's coupling $J_H$. Being the most correlated
  orbital, the DOS of the Fe $3d_{xy}$ orbital at the Fermi level is
  almost gapped for $U=6$~eV and $J_H=1.2$~eV.}
\label{fig:xy_UJ_dep}
\end{figure*}

In order to investigate the effects of $U$ and $J_H$ more explicitly
and to determine the nature of the peak/shoulder at $[-1.5,-1]$~eV we
performed calculations for different interaction parameters for the
most correlated case, the $a$-axis stretched {\cs} system.  For
computational efficiency these calculations were done at higher
temperature $\beta=40$~eV$^{-1}$.  While the height of the
quasiparticle peak is reduced at higher temperatures, the behavior of
the spectral function at $[-2,-1]$~eV is quite robust.  We considered
on-site Coulomb values $U=4$, $5$ and $6$~eV and Hund's couplings of
$J_H=0.8$, $1.0$ and $1.2$~eV.  In
Fig.~\ref{fig:z2_UJ_dep} and \ref{fig:xy_UJ_dep} we show the spectral
function $A(\omega)$ for Fe $3d_{z^2}$ and $3d_{xy}$.  An increase of
$U$ from $4$~eV to $6$~eV at a fixed $J_H$ implies only moderate
changes in the spectral function in general for all Fe $3d$ orbitals.
The Hubbard-like shoulder at $-1.2$~eV becomes more pronounced for
larger $U$ values and its maximum moves only very slightly to negative
energies ($-1.4$~eV). Due to particle-hole asymmetry, we obtain a
quasiparticle like peak slightly shifted away from the Fermi level.
On the other hand, an increase in the Hund's coupling $J_H$ for fixed
$U$ immediately renders the system very incoherent, with a strong
increase in the scattering rate and a reduction of the quasiparticle
lifetime, leading to a strong suppression of DOS at the Fermi level
and a significant shift of spectral weight to lower energies, forming
a broad lower Hubbard band located between $-4$ and $-5$~eV.  Finally,
the combined effect of $U$ and $J_H$ yields an even more well defined
lower Hubbard-like band at around $-4.5$~eV. While this characteristic
dependence on $U$ and $J_H$ is very similar for all Fe $3d$ orbitals,
we still find a strong orbital selection regarding the remaining
spectral weight at the Fermi level. Especially the Fe $3d_{xy}$
orbital is almost gapped at the Fermi level for the largest
interaction values considered (see Fig.~\ref{fig:xy_UJ_dep}), while
the least correlated Fe $3d_{x^2-y^2}$ orbital experiences basically
no suppression of the DOS at the Fermi level regardless of the
interaction parameters and instead retains a well defined
quasiparticle peak for higher values of $U$ and $J_H$ (see
Appendix~\ref{appendix:dos_other_orbs}).
We also checked the case of negligible Hund's coupling by setting $J_H=0$, 
which recovered the coherence properties even at $T=300$~K, with low effective
masses around 1.4 and a spectral function that resembled quite well the DFT
density of states.

Our results confirm the general picture of the iron pnictides being
"Hund's metals" with strong orbital separation, especially for the
strongly correlated hole-doped end systems considered in this study.
In this case, a slight increase of $J_H$ renders the system much more
incoherent and "bad metal"-like for the same value of $U$, while the
spectral weight at the Fermi level differs strongly between the
orbitals but remains finite even for larger values of $U$.

\section{CONCLUSIONS}
\label{sec:conclusion}

From our analysis of the electronic properties within LDA+DMFT in a
wide range of binding energies, we conclude that along the
isoelectronic doping series $A$Fe$_2$As$_2$ ($A={\rm K}$, Rb, Cs) as
well as the fictitious {\fr} and $a$-axis stretched {\cs}, which shows
a monotonous increase of the $a$ lattice parameters and a decrease of
the As $z$ height, correlation and incoherence of the Fe $3d$ orbitals
increase, albeit orbitally selective, and the systems show clear
features of a Hund's metal.  In this case the Hund's coupling plays
the major role and renders these materials much more incoherent than
expected from the value of the Coulomb repulsion $U$ alone. While the
most correlated orbitals ($d_{xy}$) show features that resemble those
of being close to an orbital selective Mott transition, specially for
$a$-stretched {\cs}, the system is quite deep in the incoherent bad
metal regime with a finite spectral weight at the Fermi level even for
$U=6$~eV and $J_H=1.2$~eV.  Experimentally, we predict that an
increase of the Fe-Fe distance in {\cs} by stretching will induce an
orbital dependent increase in correlations and incoherence of the
Fe $3d$ orbitals, where the Fe $3d_{z^2}$ and Fe $3d_{xy}$ orbitals
are strongly but not fully localized and the other Fe $3d$ orbitals
retain a bad metallic behaviour.  From our results we estimate the
coherence temperature  to be  located around
$50$~K in {\kf} and even lower for {\rb} and {\cs} in qualitative agreement
with the experimental observations.  These
features make the hole doped end systems of the 122 iron pnictides,
namely {\kf}, {\rb} and especially {\cs} and $a$-axis stretched {\cs} a valuable test bed to study
the behavior of strongly correlated Hund's metals and orbital-selective bad metallicity and its interplay with superconductivity.

\acknowledgments 

The authors would like to thank Felix Eilers, Kai Grube, Fr\'ed\'eric
Hardy, Christoph Meingast, Leni Bascones, Bernd B\"uchner, Stefan-Ludwig Drechsler 
and Aaram J. Kim for fruitful discussions and
gratefully acknowledge the Deutsche Forschungsgemeinschaft for
financial support through grant SPP 1458.
\begin{appendix}

\section{METHODS}
\label{appendix:methods}

\begin{figure}[t]
\includegraphics[width=0.75\linewidth]{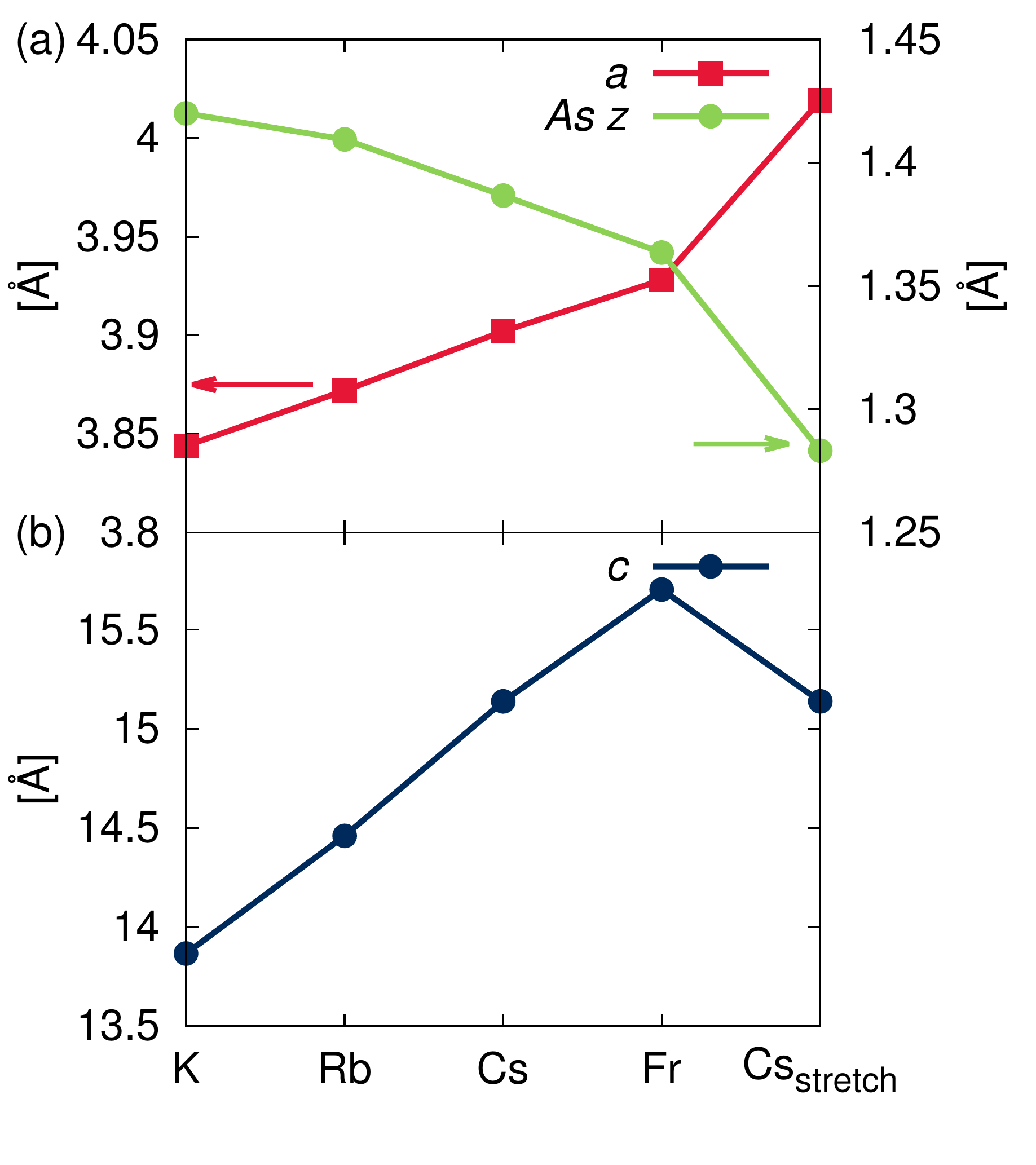}
\caption{(Color online) The structural parameters for the considered materials
{\kf} (K), {\rb} (Rb), {\cs} (Cs) (from Ref.~\onlinecite{struct_ref}), 
and the fictitious systems {\fr} (Fr) and stretched {\cs} (Cs$_{stretch}$).
(a) The $a$-parameter and absolute height of the As-atom above the Fe plane.
(b) The $c$-lattice parameter. }
\label{fig:struct}
\end{figure}

For our fully charge self-consistent LDA+DMFT calculations (see Ref.~\onlinecite{backes2014} for
detailed explanation) we use the structural parameters from
Ref.~\onlinecite{struct_ref} for the tetragonal structures of
{\kf}, {\rb} and {\cs} at room temperature.  Due to the almost perfect
linear dependence of the lattice parameters as a function of atomic radius,
we further use linear extrapolation to obtain structural parameters for
fictitious {\fr}, avoiding possible ambiguities from DFT-based relaxation methods which
do not work satisfactorily for these systems. Additionally, we prepare a structure for {\cs} that
is extended along the $a/b$-axis by $3\%$ and has a reduced relative
As $z$ height of $2\%$ to mimic a small expansion of the lattice. The
expansion is performed in both the $x$- and $y$-direction so that
tetragonal symmetry is preserved.  The lattice parameters and As $z$
position are shown in Fig.~\ref{fig:struct}.

For the DFT calculations we used the \textsc{WIEN2k}~\cite{Blaha01}
implementation of the full-potential linear augmented plane wave
(FLAPW) method in the local density approximation. The Kohn-Sham
equations were solved on 726 $k$-points in the irreducible Brillouin
zone, resulting in a $21\times 21\times 21$ $k$ mesh in the
conventional Brillouin zone.  A local orbital basis was obtained by a
projection of the Bloch wave functions to the localized Fe $3d$
orbitals, using our implementation of the method described in
Refs.~\onlinecite{Aichhorn09,Ferber2014}.  Please note that a
coordinate system which is rotated by $45^\circ$ around the $z$-axis
with respect to the conventional $I\,4/mmm$ unit cell is used. Thus,
the $x$- and $y$-axis point towards neighboring Fe atoms.  The energy
window for the bands to be considered for projection was chosen to be
$[-6, 13]$~eV, with the lower boundary lying in a gap in the density
of states (DOS). Consequently, $35$ bands on average were taken into
account for the projection, resulting in a representation of the
$k$-dependent and local non-interacting spectral function for each
orbital that is indistinguishable from the DFT result in the chosen
energy window.

The DMFT impurity problem was solved with the continuous-time quantum
Monte Carlo method in the hybridization expansion~\cite{Werner06} as
implemented in the ALPS~\cite{ALPS11,Gull11a} project.  In the
calculations we used an inverse temperature of $\beta =
80\,\mathrm{eV^{-1}}$, corresponding to the temperature of $145$ K,
unless stated differently.  A total number of at least $50\times
10^{6}$ Monte-Carlo sweeps were performed for each solution of the
impurity model and up to $90\times 10^{6}$ sweeps for the larger
interaction parameters.  For the double counting correction we used
the nominal double counting\cite{HauleCovaleny,HauleExactDC}, which
has been shown to yield significantly better agreement with
photoemission experiments\cite{HauleExactDC}, especially for low and
high binding energies, while other methods like the
FLL\cite{Anisimov93, Dudarev98} double counting scheme overestimate
the valence charge and underestimate a possible Mott
gap\cite{HauleExactDC}.
The interaction parameters were used in the
definition of the Slater integrals~\cite{Liechtenstein95} $F^k$ with
$U=F^0$ and $J_H=(F^2+F^4)/14$. For the on-site Coulomb interaction we
considered a value of $U=4\,\mathrm{eV}$ and for Hund's rule coupling
$J_H=0.8\,\mathrm{eV}$, unless stated differently.  
We calculate the effective masses directly from the impurity self-energy via
\begin{equation}
    \frac{m^\ast}{m_\mathrm{LDA}} = 1 - \left.\frac{\partial\mathrm{Im}\Sigma(i\omega)}{\partial \omega}\right|_{\omega\rightarrow 0^+},
    \label{eq:meff}
\end{equation}
with the quasiparticle weight being defined as the inverse of the effective mass $Z=\left[\frac{m^\ast}{m_\mathrm{LDA}}\right]^{-1}$.
The continuation of the Monte Carlo data to the real axis was done by stochastic
analytic continuation\cite{beach2004}.

\begin{figure}[t]
\includegraphics[width=1\linewidth]{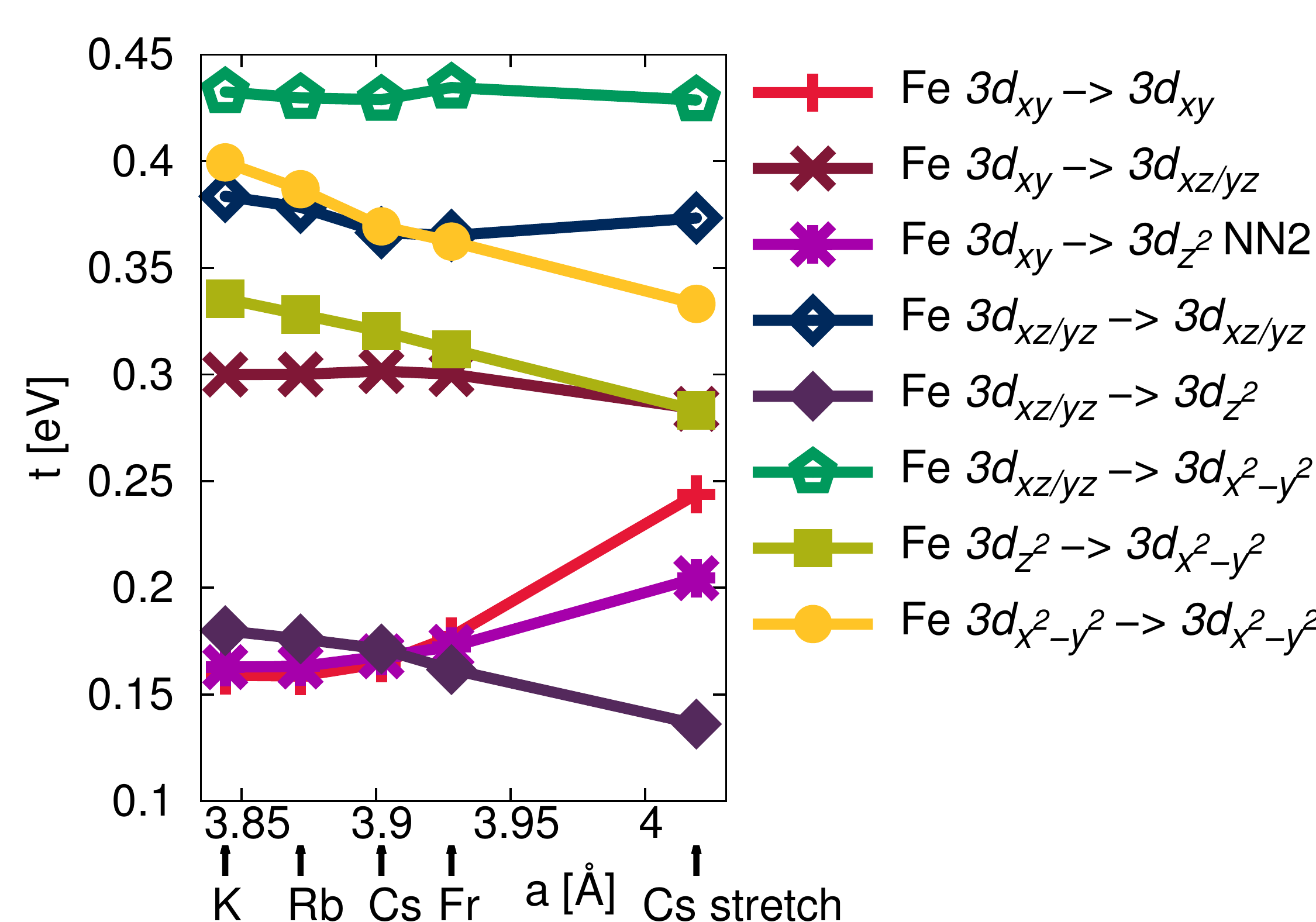}
\caption{(Color online) Eight largest effective Fe-Fe hopping
  parameters $t$ obtained from a 10-band tight
  binding fit. Also shown are some second nearest
  neighbor parameters indicated by the label NN2.  The non-trivial evolution
  of the hopping values with increasing lattice parameter and reduced As$_z$ height leads to
  different degrees of localization and effects of correlation in the
  Fe $3d$ orbitals along the series {\kf}, {\rb}, {\cs} and {\fr}.}
\label{fig:hoppings}
\end{figure}

For obtaining the hopping matrix elements we obtained a tight-binding
Hamiltonian from projective Wannier functions~\cite{FPLOtightbinding}
from DFT, generated by the all-electron full-potential local orbital
(FPLO)\cite{FPLO} code, using a 10 (16) orbital model, including the
Fe $3d$ only (10-orbital model)~\cite{bascones2009} or Fe $3d$ and As $4p$ orbitals
(16-orbital model).

\section{10-band tight binding model}
\label{appendix:10band_tb}

In order to quantify at the level of DFT the effects of negative
pressure introduced by isovalent doping in $A$Fe$_2$As$_2$ ($A={\rm
  K}$, Rb, Cs, Fr), we calculated the Fe-Fe hopping matrix elements
via projective Wannier functions. The absolute values of the hopping
parameters are plotted in Fig.~\ref{fig:hoppings}. There are two main
contributions that affect the values of the Fe-Fe hopping parameters:
First, the increase of the interatomic distances due to increasing
atomic radii of the alkali ions implies a decrease of the direct Fe-Fe
hopping. Second, due to elongation of the Fe-As tetrahedron, the As
atom moves closer to the Fe-Fe plane. This reduction of the As$_z$
height leads to an increase of the indirect hopping along the path
Fe-As-Fe.  The total contribution of these two effects translates into
a non-trivial behavior of the Fe-Fe effective hoppings along the
doping series (K,Rb,Cs)Fe$_2$As$_2$.  The Fe $3d_{xy}$- Fe $3d_{xy}$
effective hopping is the smallest in KFe$_2$As$_2$ due to the almost
perfect cancellation of the two contributions as pointed out in
Ref.~\onlinecite{haulekotliar2011}.  As the lattice parameter
increases, the indirect Fe-As-Fe hopping contribution outweighs the
contribution coming from the direct Fe-Fe orbital overlap. This causes
a slight increase of the hopping parameters from {\kf} to {\cs}.
These two contributions are very similar for hoppings between Fe
$3d_{xz/yz}$ orbitals.  The trend in the hoppings between Fe
$3d_{z^2}$ and $3d_{x^2-y^2}$ orbitals is less affected by changes of
the indirect hopping contribution and shows a small overall decrease
in the hopping to the neighboring Fe $3d$ orbitals.

\section{16-band tight binding model}
\label{appendix:16band_tb}
\begin{figure}[t]
\includegraphics[width=1\linewidth]{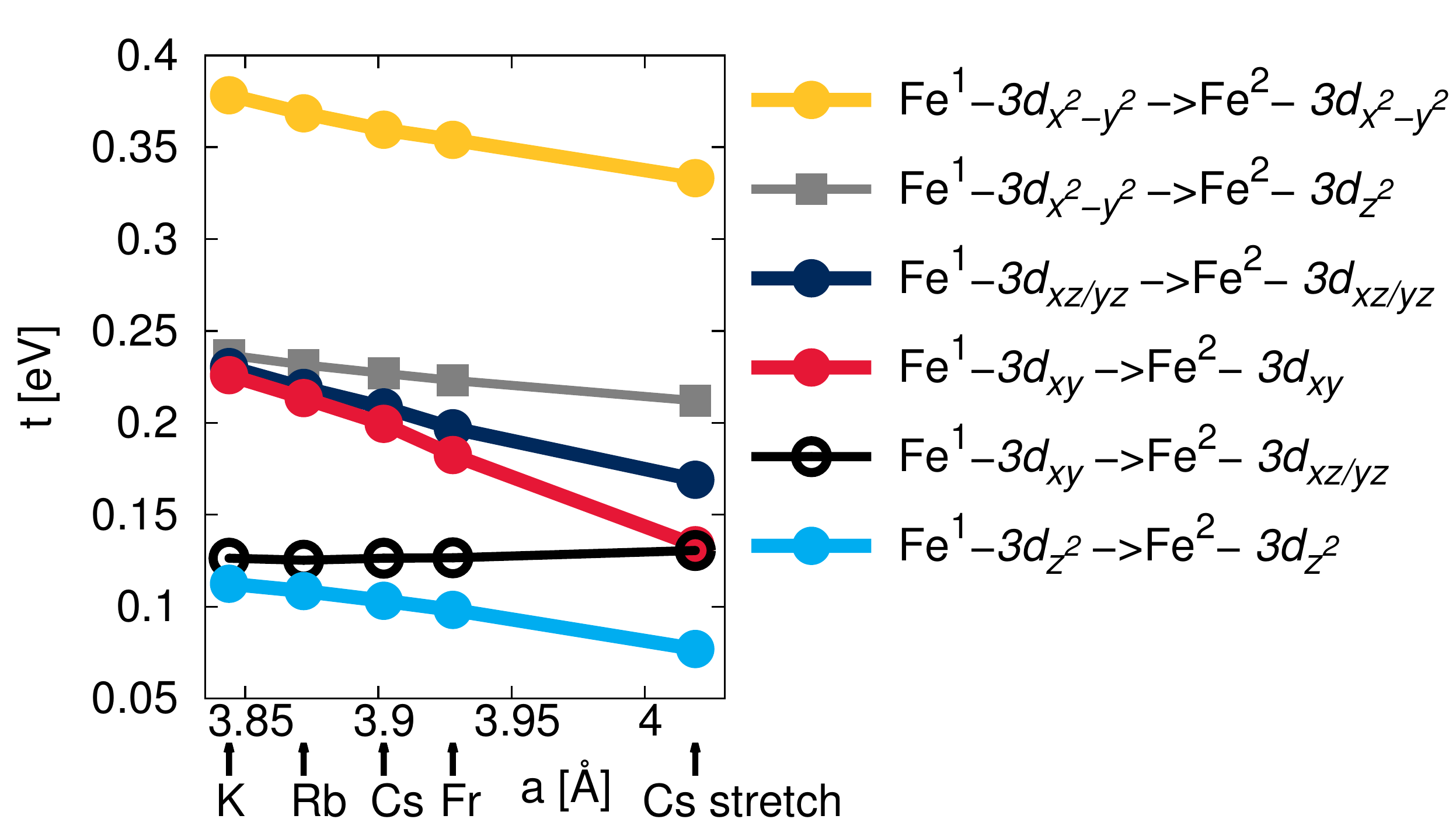}
\caption{(Color online) The first six largest hopping parameters $t$
  for the Fe-Fe hopping obtained from a 16-band tight binding
  fit. Fe$^i$ denotes the $i$-th atom out of the two equivalent iron
  atoms in the irreducible Brillouin zone.  }
\label{fig:hoppings_16b}
\end{figure}

In Fig.~\ref{fig:hoppings_16b} we show the six largest Fe-Fe hopping
parameters obtained from a 16-band tight binding fit, encompassing the
Fe $3d$ and As $4p$ orbitals. The overall monotonous decrease
resembles the increase of the interatomic distance that leads to a
reduced overlap of the neighboring Fe $3d$ orbitals. As noted in the
main text, the indirect hopping through the As $4p$ orbitals has an
important effect on the effective hopping parameters.  Taking only the
direct Fe-Fe hopping into account, we observe the expected decrease of
the hopping parameters which resembles the reduced hybridization as
the lattice parameters are increased.  In combination with the
indirect hopping via the As atom, this leads to a nontrivial behaviour
of the effective hopping parameters.

\section{DOS for other orbitals}
\label{appendix:dos_other_orbs}

\begin{figure}[t]
\includegraphics[width=1\linewidth]{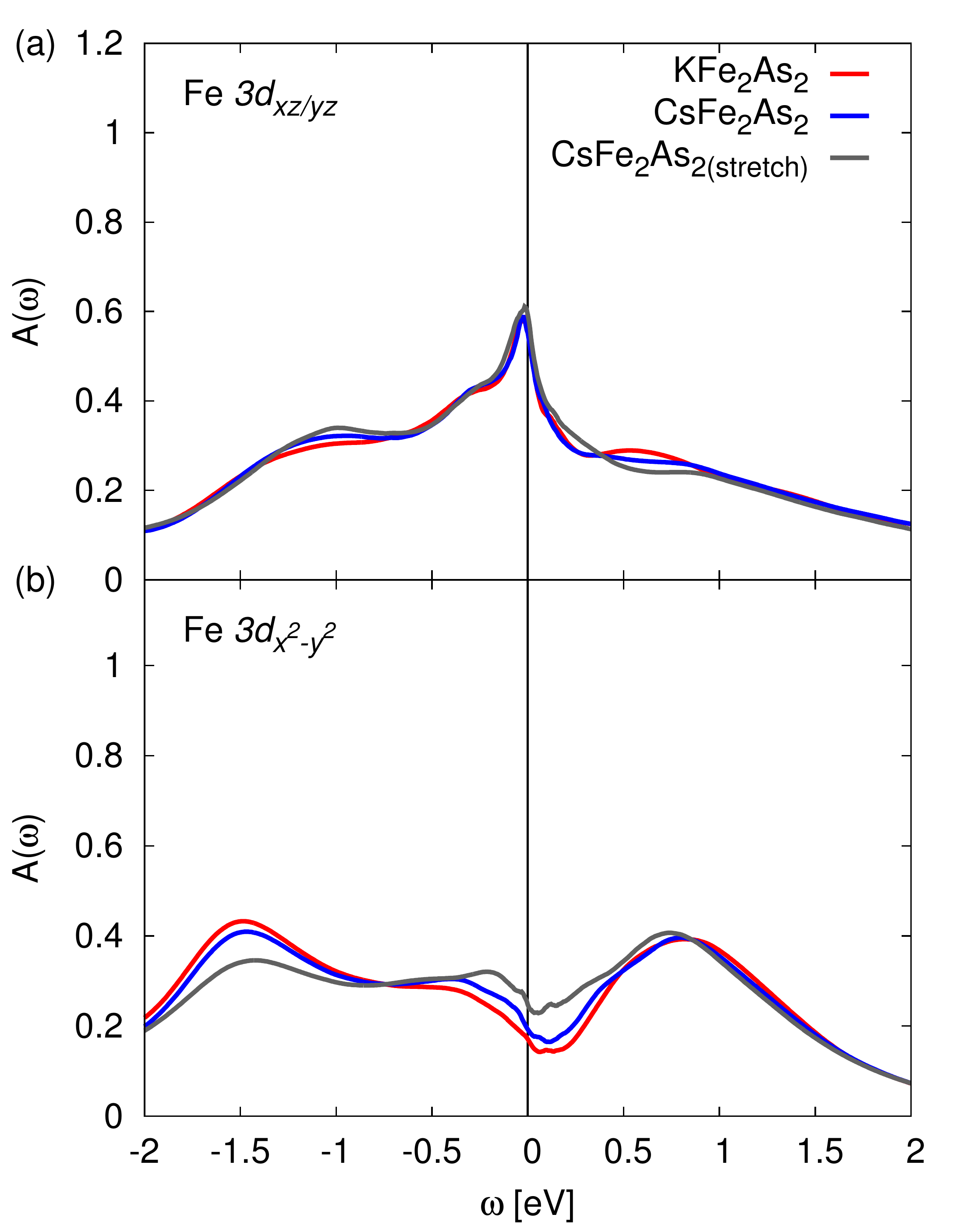}
\caption{(Color online) The density of states for (a) the Fe $3d_{xy}$
  orbital and (b) the Fe $3d_{x^2-y^2}$ orbital for $U=4$~eV and
$J_H=0.8$~eV as obtained from
  LDA+DMFT for the three compounds {\kf}, {\cs} and an expanded
  structure of {\cs}.}
\label{fig:xz_z2_doslarge}
\end{figure}

\begin{figure*}[t]
\includegraphics[width=1\linewidth]{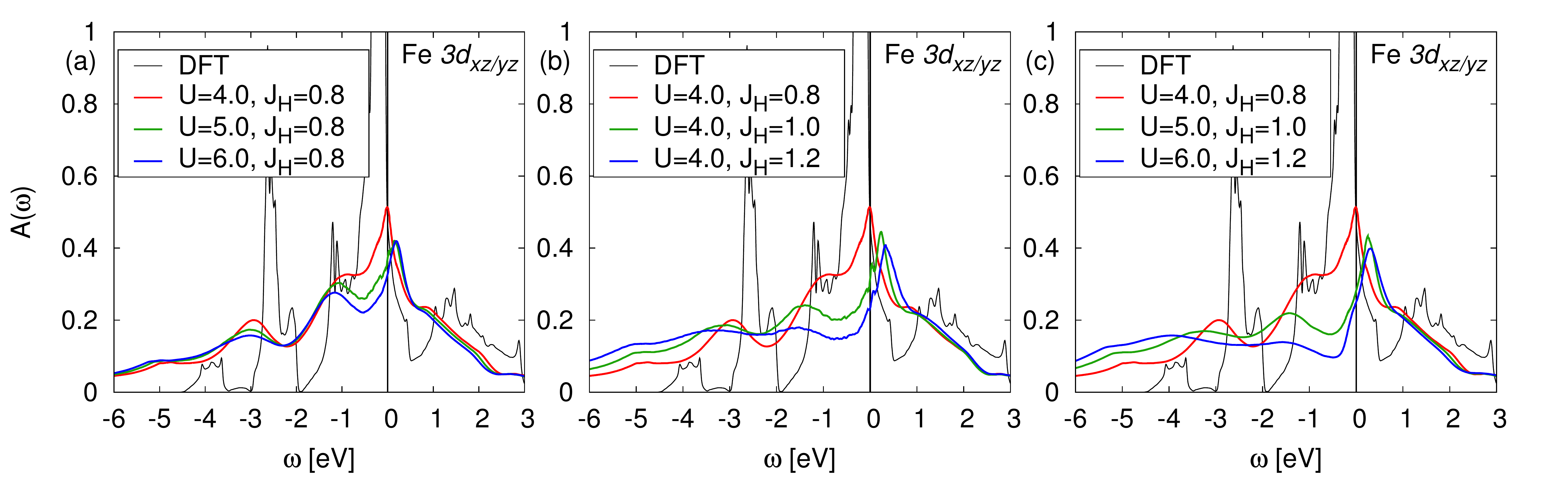}
\includegraphics[width=1\linewidth]{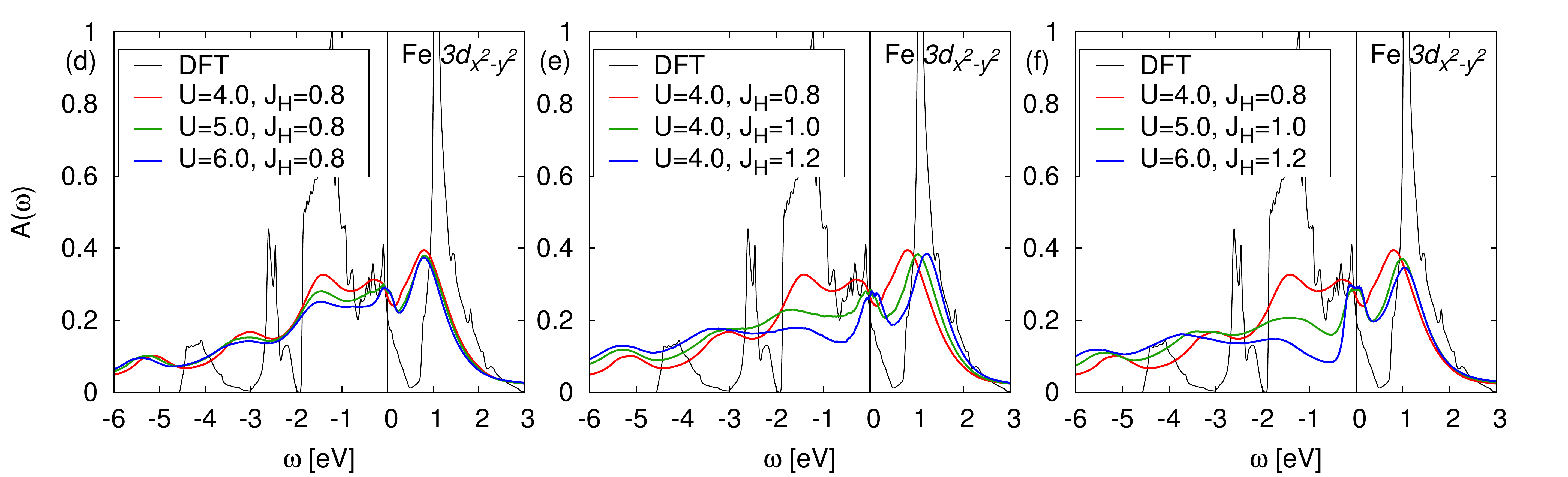}
\caption{(Color online) The density of states for the Fe $3d_{xz/yz}$
  and $3d_{x^2-y^2}$ orbital of the stretched {\cs} compound as a
  function of the on-site Coulomb repulsion $U$ and Hund's coupling
  $J_H$. (a)+(d) An increase only in $U$ leads to a slightly better
  pronounced lower Hubbard band-like feature but otherwise no
  qualitative changes to the high energy features.  (b)+(e) The Hund's
  coupling $J_H$ greatly increases the decoherence of the electronic
  states at the Fermi level and leads to a significant shift of
  spectral weight down to negative energies. (c)+(f) The combined effect of
  $U$ and $J_H$ is qualitatively very similar to an increase in $J_H$
  alone.}
\label{fig:xzx2y2_UJ_dep}
\end{figure*}

In Fig.~\ref{fig:xz_z2_doslarge} we show the local spectral function
of the Fe $3d_{xz/yz}$ and $3d_{x^2-y^2}$ orbital for {\kf}, {\cs}
and $a$-axis stretched {\cs}. Compared to the other Fe $3d$ orbitals they are
less affected by an increase of the lattice parameter $a$. Similar to
the $3d_{z^2}$ and $3d_{xy}$ orbital a small Hubbard-like peak becomes
more pronounced in the $3d_{xz/yz}$ orbital, while the $3d_{x^2-y^2}$
orbital shows the opposite trend, increasing its spectral function at
the Fermi level at the cost of decreasing it at negative energies.

In Fig.~\ref{fig:xzx2y2_UJ_dep} we show the dependence of the Fe
$3d_{xz/yz}$ and $3d_{x^2-y^2}$ orbital spectral function on $U$ and
$J_H$. The results are very similar to the other orbitals, with the
effect of increasing $U$ being much less extreme than that of
$J_H$. While at higher $U$ the spectral functions still mimics the LDA
result, an increase in $J_H$ results in a large shift of spectral
weight to negative energies. The $3d_{x^2-y^2}$ orbital shows the smallest
degree of correlations and is the only orbital that retains a well 
defined quasiparticle peak at the Fermi
level even for larger interaction parameters.

\end{appendix}

\end{document}